\documentclass[5p, preprint,twocolumn, 9.5pt]{elsarticle}

\usepackage{lineno}
\usepackage[hypertexnames=false,
   colorlinks,linkcolor={blue},citecolor={blue},urlcolor={blue},
   pdfstartview={FitV},unicode,breaklinks=true]{hyperref}
\usepackage{amssymb}
\usepackage{graphicx}
\usepackage{caption}
\usepackage{subcaption}
\usepackage{amsmath}
\usepackage{amssymb,latexsym}
\usepackage{savesym}
\savesymbol{AND}
\usepackage{algorithmic}
\usepackage{algorithm}
\usepackage{color}
\usepackage{xcolor}
\usepackage{multirow}
\usepackage{vector}

\newcommand \mc[1]{\mathcal{#1}}
\newcommand \cO{\mathcal{O}}
\newcommand \cMA{\mathcal{MA}}
\newcommand \cMAT{\mathcal{MAT}}
\newcommand \cM{\mathcal{M}}
\newcommand \mR{\mathbb{R}}

\journal{Graphical Models}

\begin{document}

\begin{frontmatter}



\title{Computing a Compact Spline Representation of \\the Medial Axis Transform of a 2D Shape}

\author[mymainaddress]{Yanshu Zhu\corref{mycorrespondingauthor}}
\cortext[mycorrespondingauthor]{Corresponding author}
\ead{yszhu@cs.hku.hk}
\author[mymainaddress]{Feng Sun}
\author[mymainaddress]{Yi-King Choi}
\author[mysecondaryaddress]{Bert J\"{u}ttler}
\author[mymainaddress]{Wenping Wang}
\address[mymainaddress]{Department of Computer Science, The University of Hong Kong, Hong Kong}
\address[mysecondaryaddress]{Institute of Applied Geometry, Johannes Kepler University, Linz, Austria}


\begin{abstract}
We present a full pipeline for computing the medial axis transform of an arbitrary 2D shape. The instability of the medial axis transform is overcome by a pruning algorithm guided by a user-defined Hausdorff distance threshold. The stable medial axis transform is then approximated by spline curves in 3D to produce a smooth and compact representation. These spline curves are computed by minimizing the approximation error between the input shape and the shape represented by the medial axis transform. Our results on various 2D shapes suggest that our method is practical and effective, and yields faithful and compact representations of medial axis transforms of 2D shapes.
\end{abstract}


\end{frontmatter}



\section{Introduction} \label{sec:intro}

The notion of the medial axis transform was first introduced by Blum~\cite{blum1967} as an intrinsic shape representation. The medial axis of an object $\cO$ is the set of interior points having at least two closest points on the boundary $\partial \cO$ of $\cO$. In the 2D space, each point on the medial axis is the center of a circle, namely a {\em medial circle}, which is the maximal inscribed circle contained in $\cO$ and tangent to $\partial \cO$ in at least two points.
To encode the complete shape information of the object, each point on the medial axis is assigned with the radius, which could be 0, of its associated medial circle.
Therefore, a radius function could be defined on the medial axis.
The medial axis coupled with a radius function is referred to as the {\em medial axis transform} (MAT). Each point in the MAT, called {\em medial point}, has three dimensions, which indicates its 2D position and the radius.
The MAT is a complete shape representation in the sense that the object boundary can be reconstructed exactly from its MAT as the envelope of all the medial circles.

The MAT encodes rich information of a shape, such as local thickness, symmetry and its part structure, which is not possessed by alternative boundary surface representations.
Therefore the MAT has been used extensively in a wide spectrum of applications, including shape analysis~\cite{bouix2005hippocampal}, shape deformation~\cite{yoshizawa2007skeleton} and artistic rendering~\cite{gooch2002artistic}. Detailed discussions on properties and applications of the MAT can be found in the book~\cite{siddiqi2008}.

The MAT, on the other hand, is well-known suffering from the {\em instability} problem:
small variations of the shape boundary may yield a large change to its MAT.
While boundary noise is ubiquitous in data acquisition due to errors introduced in scanning, sampling and other numerical processing, the medial axis thus often has excessive geometric complexity and pathological topology, rendering it generally useless in practice unless it is cleaned up.
A lot of existing algorithms have been developed to resolve the instability issue.
As a common practice, unstable branches of the MAT induced by boundary noise are pruned based on certain measures~\cite{chazal2005lambda,giesen2009scale,Sud:2005}.
Different criteria have been introduced to characterize the difference between the original shape and the reconstructed shape, 
or describe the remaining stable MAT with some intrinsic measures~\cite{attali2009stability}.
These methods focus on simplifying the topology of the MAT by pruning unstable branches,
producing a topologically clean MAT which is nevertheless still represented by a large number of sample medial points.
This is partially due to the prevailing choice taking the union of the sample medial circles (or medial spheres in 3D) as an approximation when evaluating the approximation error during both pruning stage and shape reconstruction stage.
As a consequence, a large number of medial circles are often necessary to attain a good approximation to the MAT~\cite{miklos2010discrete,stolpner2012medial}.

\begin{figure}
\centering
    \begin{subfigure}[t]{0.45\linewidth}
    \centering
    \includegraphics[width=\linewidth]{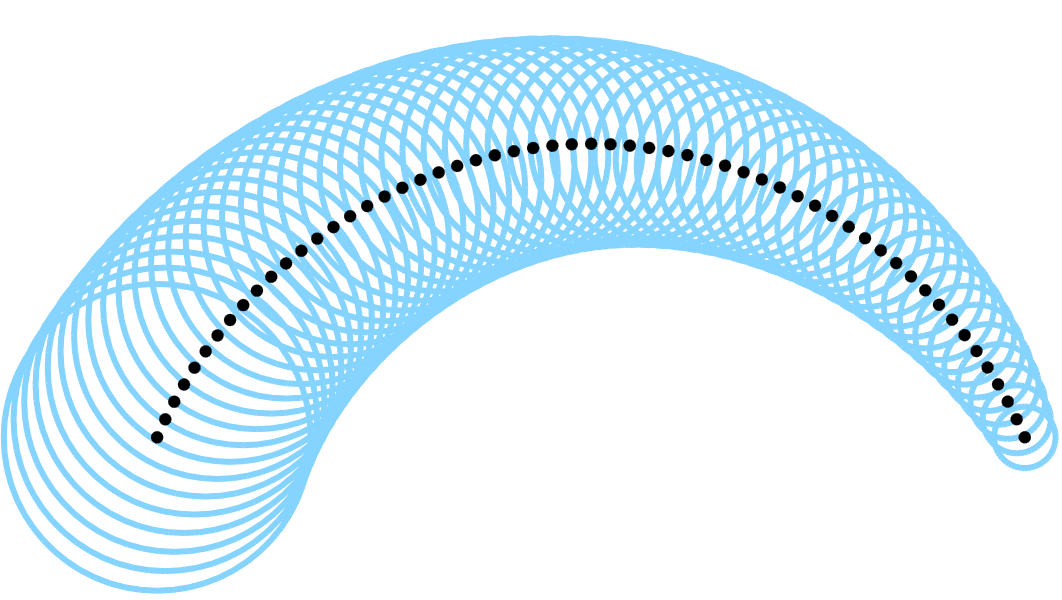}
     \caption{}
    \label{fig:recon:uob}
    \end{subfigure}
    \hfill
    \begin{subfigure}[t]{0.45\linewidth}
    \centering
    \includegraphics[width=\linewidth]{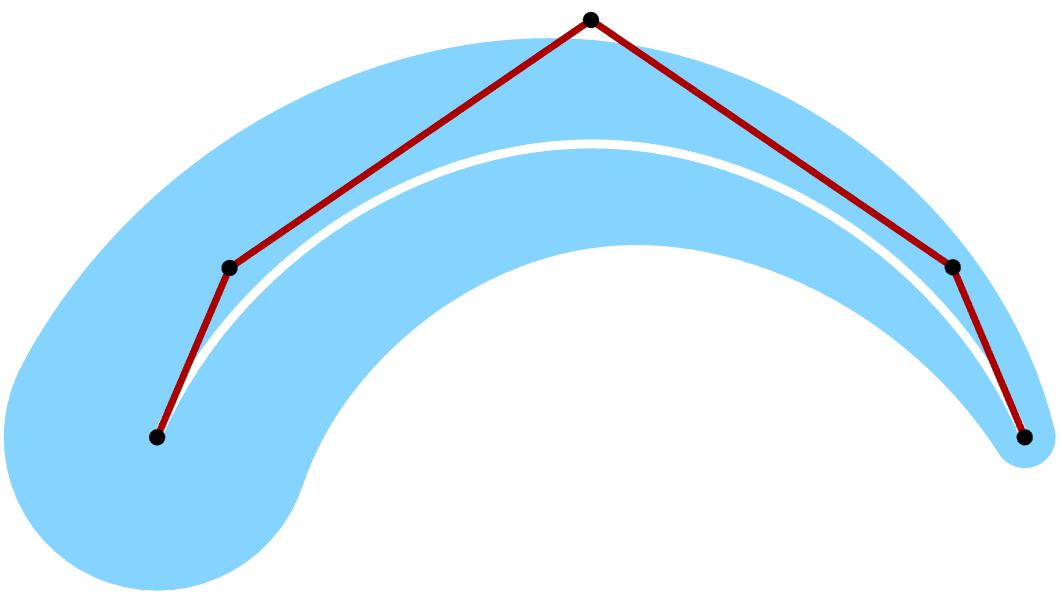}
    \caption{}
    \label{fig:recon:curve}
    \end{subfigure}
\caption{\small Given a shape swept by a moving circle, its medial axis transform is approximated as a sample set of 57 medial circles. (a) The approximate shape is the union of the medial circles. (b) On the other hand, to achieve the same accuracy with spline medial axis approximation, we need only $5$ control points. The control polygon of the spline curve is shown in red. }
\label{fig:recon}
\end{figure}

To achieve a smooth representation and further reduce the geometric complexity of the MAT, we propose to represent the MAT as spline curves. See Fig.~\ref{fig:recon} for a comparison of the two shape representations. The medial axis also possesses piecewise $C^2$ continuity in each medial branch. Although a similar smooth representation of the MAT has been used for modeling and segmentation purposes~\cite{yushkevich2003continuous}, a fully automatic way of obtaining a smooth MAT for an arbitrary shape is still missing.

In this work, we propose a complete pipeline which automatically computes a {\em stable and compact} medial axis transform which accurately approximates an arbitrary 2D shape.
Given an error threshold $\hat \varepsilon$, our algorithm guarantees that the Hausdorff distance between the boundary of the original shape and the boundary of the reconstructed shape is at most $\hat \varepsilon$.
Our method involves the pruning of unstable medial branches with an error-driven filtering process and the computation of a compact and accurate spline approximation to the MAT.

Compared to other works on medial axis computation, our algorithm possesses the following advantages:
\begin{itemize}
  \item \textbf{Topological filtering with error control}: Our filtering process is guided by a user-defined error threshold $\hat \varepsilon$ to ensure approximation accuracy while removing noisy, unstable branches as much as possible.
  \item \textbf{Computing a compact geometric representation}: We use spline curves to approximate the MAT, resulting in a highly compact representation. An optimization process is developed to make sure that the reconstructed boundary best fits the input shape, meeting a user specified error tolerance.
\end{itemize}

This paper is organized as follows. We start with a brief review of the previous work related to medial axis computation in Section~\ref{sec:related}.  We then define the piecewise smooth medial representation in Section~\ref{sec:medialrep}.  In Section~\ref{sec:compframe}, we introduce our main algorithm; the implementation details are then provided in Section~\ref{sec:implementation}.  We present experimental results in Section~\ref{sec:experiments} and finally conclude the paper in Section~\ref{sec:conclusion}.


\section{Related Work} \label{sec:related}

There is a vast amount of research studies about medial axis computation and representation. Here, we will review only those which are in close relation to our work.

\subsection{Medial axis computation}\label{sect:voronoi medial axis}

Exact medial axis computation is possible only for simple or special shapes, such as polyhedra~\cite{culver2004,etzion1999}.
For free-form shapes, medial axis approximations are widely used in practice. There are several main approaches to computing the medial axis approximation: pixel or voxel-based methods that compute the medial axis using a thinning operation~\cite{lam1992thinning}; methods based on distance transform~\cite{foskey2003efficient,hirota2000fast,Kimmel1995,siddiqi2002hamilton}, often performed on a regular or adaptive grid;
the divide-and-conquer methods~\cite{aichholzer2009medial}, performed on spline curve boundaries;
the tracing approaches~\cite{ramanathan2003constructing}, by tracing along the shape boundary or the seam curves;
and the Voronoi diagram (VD) based methods~\cite{chazal2005lambda,amenta1999surface,amenta2001power,attali1997computing,tam2003shape}.

\begin{figure}[htbp]
\centering
    \begin{subfigure}[t]{0.30\linewidth}
    \centering
    \includegraphics[width=\linewidth]{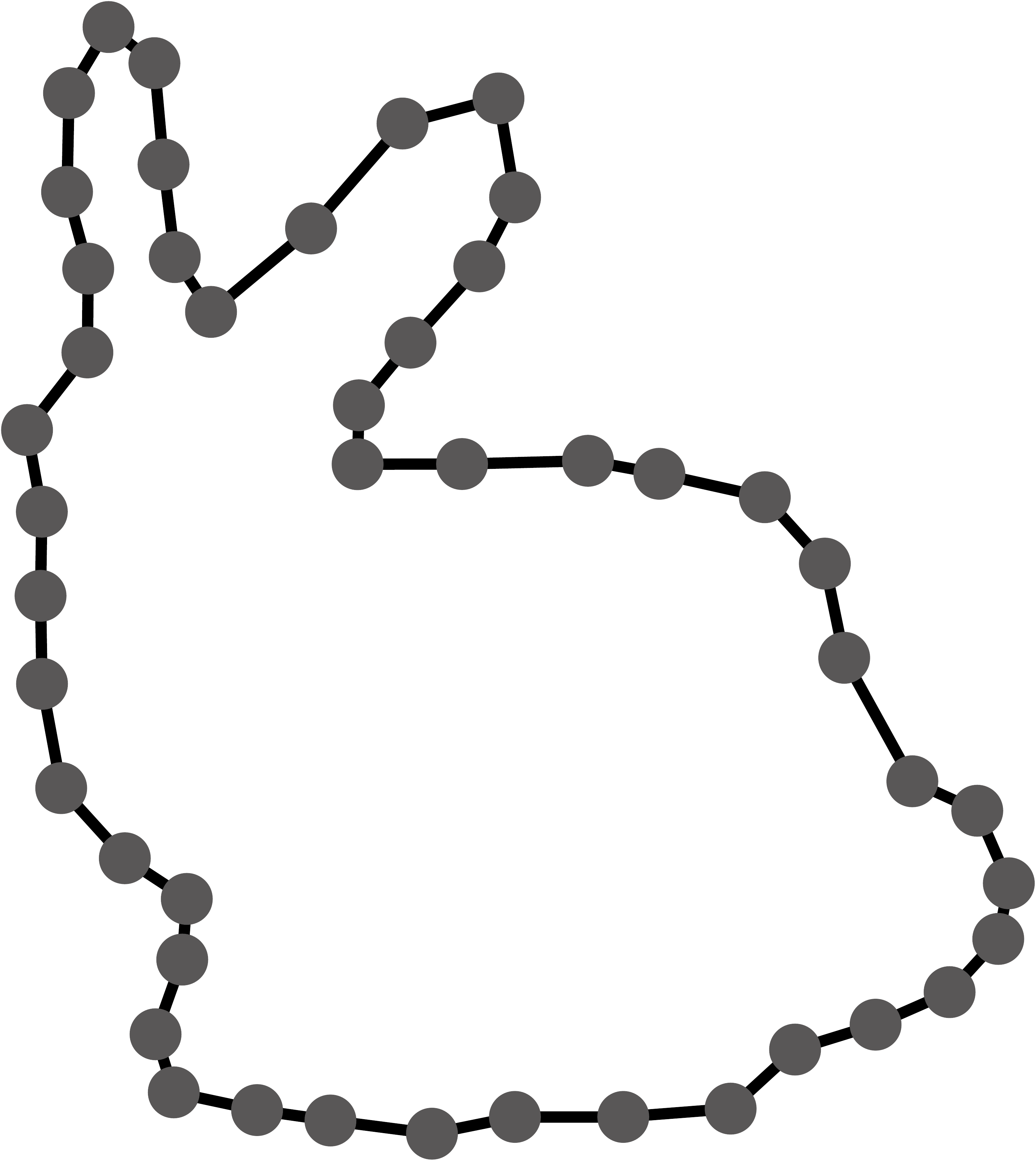}
    \caption{}
    \label{fig:rabbit:contour}
    \end{subfigure}
    \begin{subfigure}[t]{0.50\linewidth}
    \centering
    \includegraphics[width=\linewidth]{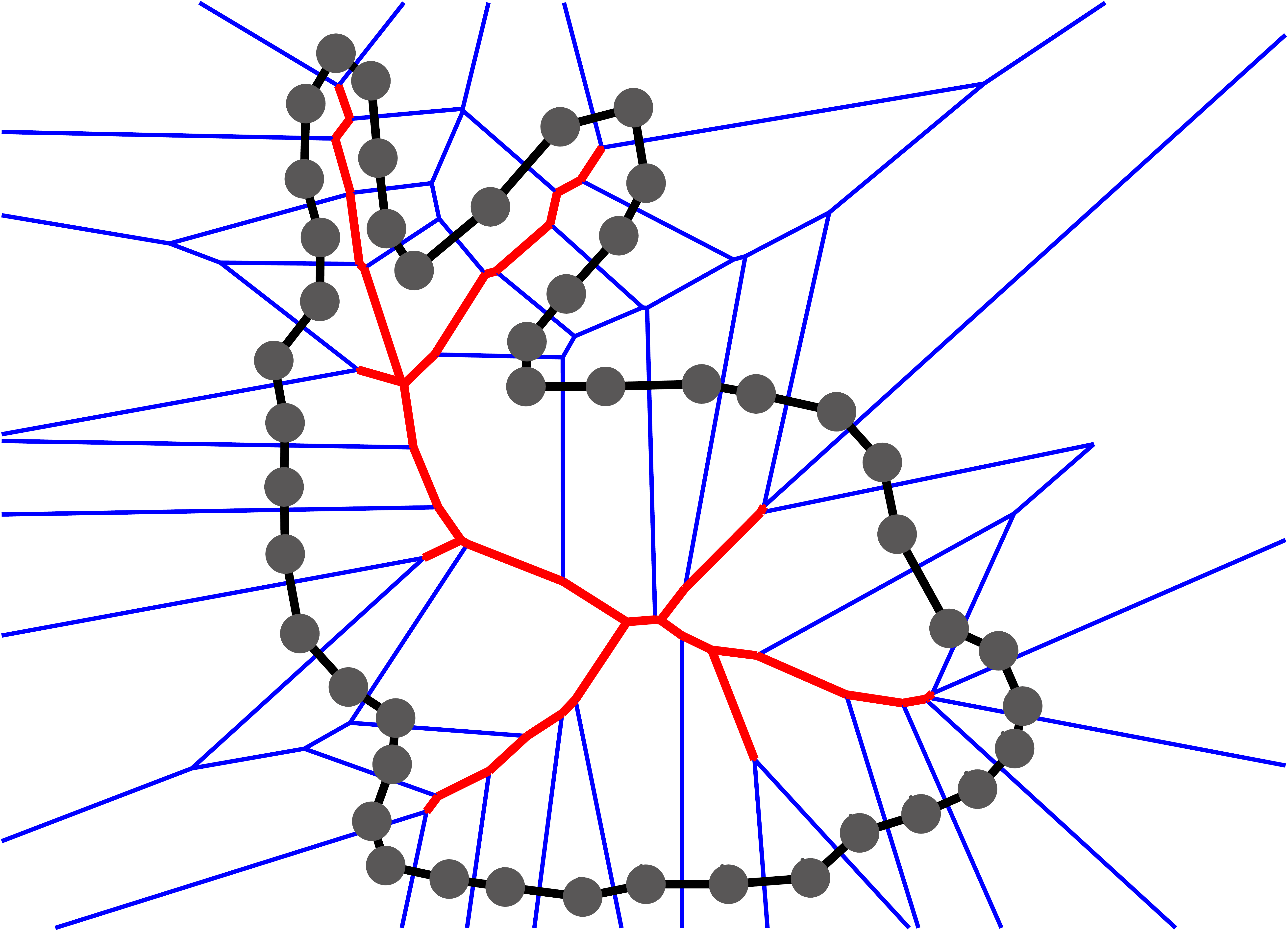}
    \caption{}
    \label{fig:rabbit:vd}
    \end{subfigure}
\caption{\small Obtaining an initial discrete medial axis transform using the Voronoi diagram based method. (a) Boundary sampling of a 2D shape. (b) The Voronoi diagram of the boundary of sampling points, with the internal part of the Voronoi diagram (in red) approximating the medial axis.}
\label{fig:rabbit}
\end{figure}

Among these, the VD based approach stands out due to its theoretical guarantee and efficient computation. As a preprocessing step, we obtain an initial discrete medial axis of a shape using the VD based algorithm. The VD based method assumes that the boundary of an input shape $\cO$ is a smooth curve and is sampled by a dense discrete set $\bvec{P}$ of points (Fig.~\ref{fig:rabbit:contour}), with the sampling density determined by the local feature size~\cite{amenta1999surface} in order to capture the boundary topology correctly.
The Voronoi diagram of $\bvec{P}$ is computed and the Voronoi vertices interior to $\cO$ are taken to approximate the medial axis of Voronoi diagram (Fig.~\ref{fig:rabbit:vd}), since a point on the Voronoi diagram is also characterized by having at least two closest points among the sample points.

\begin{figure*}
\centering
    \begin{subfigure}[t]{0.32\linewidth}
    \centering
    \includegraphics[width=\linewidth]{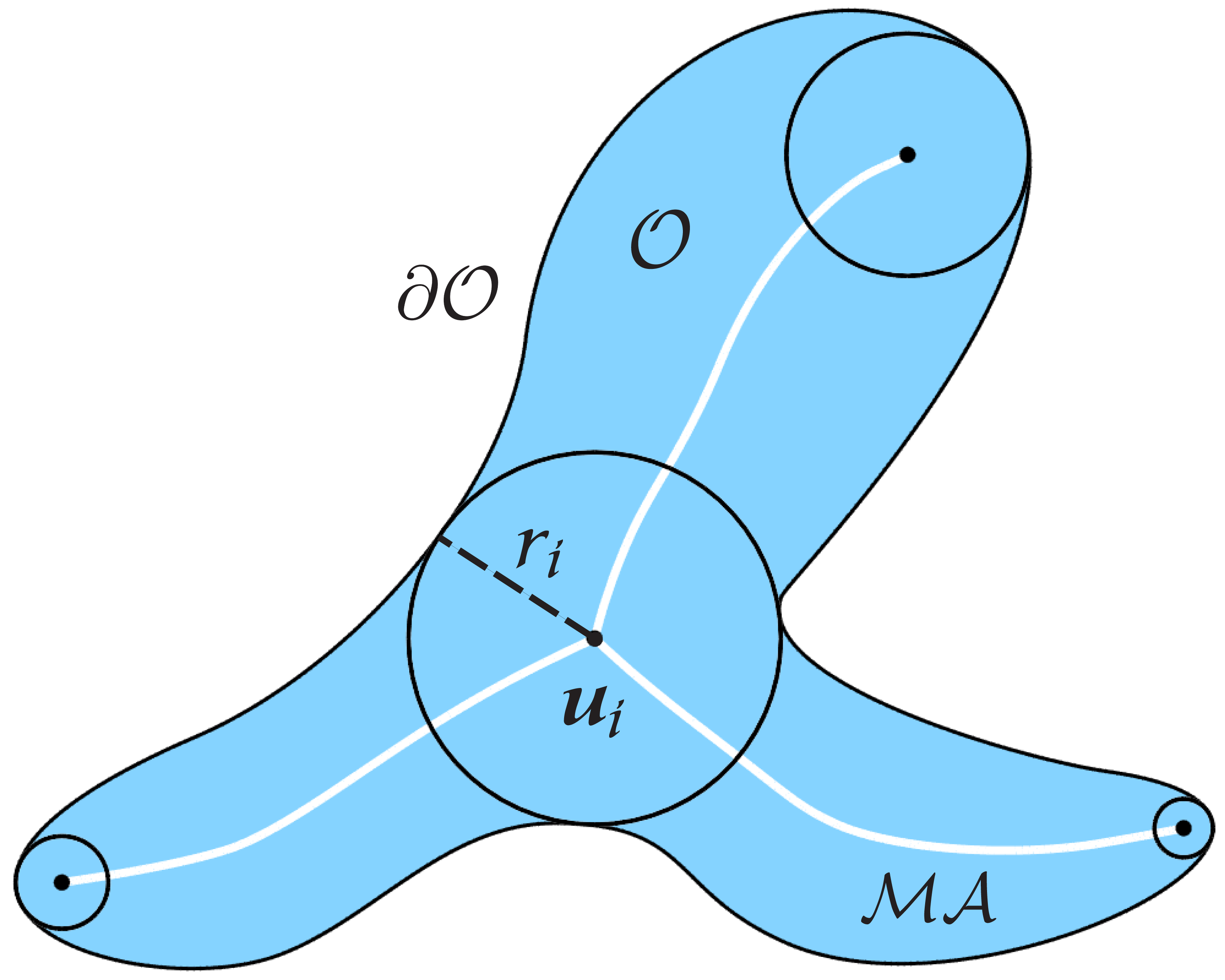}
    \caption{}
    \label{fig:branchjoint:a}
    \end{subfigure}
    \hfill
    \begin{subfigure}[t]{0.32\linewidth}
    \centering
    \includegraphics[width=\linewidth]{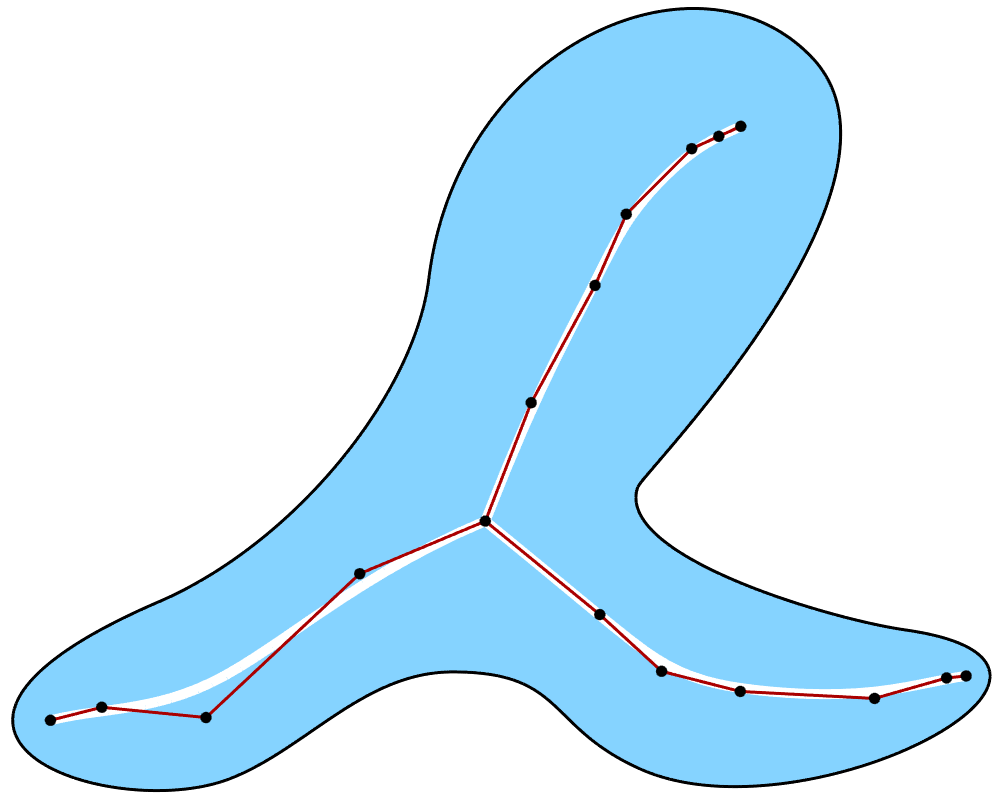}
    \caption{}
    \label{fig:branchjoint:b}
    \end{subfigure}
    \hfill
    \begin{subfigure}[t]{0.32\linewidth}
    \centering
    \includegraphics[width=\linewidth]{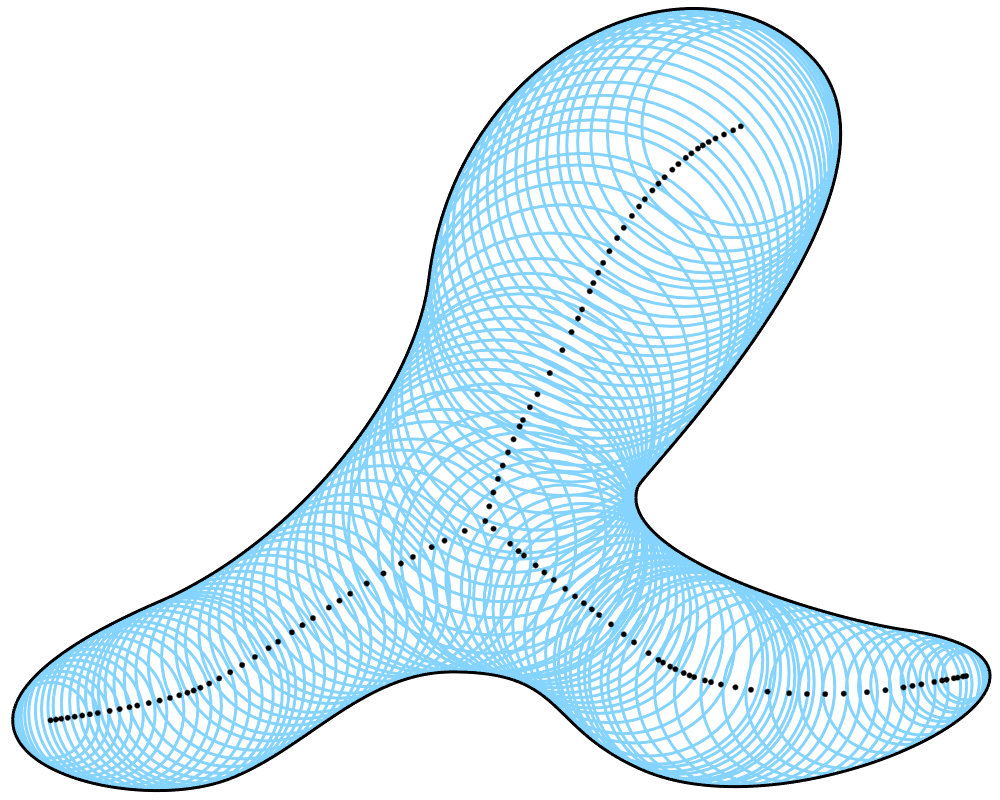}
    \caption{}
    \label{fig:branchjoint:c}
    \end{subfigure}
    \hfill
\caption{\small Illustration of the medial axis transform. (a) A medial axis consisting of three branches connected at a joint. (b) The branches are represented by cubic B-spline curves, shown with their control polygons, defined by 17 control points. (b) The reconstructed shape from the medial axis transform can be obtained from our envelope reconstruction of a densely sampled medial axis and its associated medial circles.
}
\label{fig:branchjoint}
\end{figure*}

\subsection{Handling instability}\label{sec:stablecomputation}

Many studies have been conducted to understand and resolve the instability problem of the MAT. We review here several typical methods, whereas a survey can be found in~\cite{attali2009stability}.
One general approach is to define certain measures for the significance of a medial point, and to filter medial points against a user-defined threshold, thereby removing unstable branches of the medial axis.
Examples include 
the angle-based methods which consider the separation angle or the object angle (i.e., the angle spanned by the closest contacting points)~\cite{foskey2003efficient,Attali1996,dey2004approximate} and scaled axis transform (SAT) which essentially exploits the rate of change of the radius function as the filtering condition~\cite{giesen2009scale}. 

Another approach of computing a stable MAT is to consider the difference between the initial shape and the
reconstructed shape from the pruned MAT~\cite{attali2009stability}.
The filtering step in our algorithm resembles this latter approach by considering the
Hausdorff distance between the boundary of the input and the approximate shape to ensure the approximation accuracy of the output stable MAT.

\subsection{Smooth medial axis representation}

Yushkevich et al.~\cite{yushkevich2003continuous} propose a continuous medial representation by modeling the MAT with cubic B-splines, as an extension of its discrete counterpart called the \emph{m-rep}~\cite{pizer1999segmentation}.
The m-rep is built upon a sparse set of medial atoms, each of which encapsulates the 2D position of a medial point and the corresponding spoke vectors from the 2D medial point to the closest points on the object boundary.
The continuous m-rep~\cite{yushkevich2003continuous} (cm-rep), on the other hand, uses control points in cubic B-splines to describe the MAT, which must meet specific constraints defined on the implied boundary.
In applying cm-reps to object modeling and image segmentation, a template cm-rep model is first built manually which is then deformed to fit a target shape.
There is currently no method for automatically computing a smooth curve approximation to the MAT of a 2D shape.

The smooth medial representation we propose in this work is inspired by the cm-rep. However, instead of medial atoms, we adopt the simple medial points as our basic control entities.  While the shape boundary is explicitly given by the trace of the spoke vectors for a cm-rep, the shape represented by our smooth medial axis transform is implicitly described by the union of the envelopes of adjacent medial circles.
Using medial circles not only provides a more compact representation but also relieves the burden of ensuring the shape boundary as defined by the spoke vectors is intact and consistent, without compromising the accuracy in shape approximation.


\section{Preliminaries}\label{sec:medialrep}

In this section we will introduce the notation for describing our method. Refer to Fig.~\ref{fig:branchjoint} for an illustration of the medial axis transform of a 2D shape.
Consider a 2D shape $\cO$ as a compact connected subset of $\mR^2$ with boundary $\partial \cO$ (Fig.~\ref{fig:branchjoint:a}). The medial axis $\cMA$ of $\cO$ is defined as the set of points inside $\cO$ with at least two nearest neighbors on $\partial \cO$. Medial circles could be described from points in $\cMA$, associated with the local thickness (radius) values. The medial axis transform $\cMAT$ of $\cO$ is the set of such medial circles.
Each point in $\cMAT$, which we call a {\em medial point}, is specified by a 3D vector $\bvec{v}_i = (\bvec{u}_i, r_i)$, where $\bvec{u}_i = (x_i, y_i) \in \mR^2$ denotes the medial position and $r_i \in \mR$ is the distance from $\bvec{u}_i$ to the shape boundary $\partial \cO$.
In other words, $\bvec{v}_i$ describes a 2D medial circle, centered at $\bvec{u}_i$ with radius $r_i$, which is tangent to $\partial \cO$ in at least two points. In the extreme case, for example, a sharp corner of $\partial \cO$, the radius is actually 0.

A medial point $\bvec{v}_i$ tangent to $\partial \cO$ at two distinct points is called a {\em regular} medial point.  A {\em branch} is a maximal curve segment comprising regular medial points.  A {\em joint} is a medial point at which three or more branches meet, and hence a joint has three or more tangent points with $\partial \cO$.  An {\em end-point} is a medial point corresponding to an oscillating circle of $\partial \cO$.

Given a 2D shape $\cO$, we consider an approximation to its $\cMAT$ by a {\em piecewise smooth medial axis transform}, denoted $\cM = \{C_j\}$ that consists of a set of {\em medial spline curves} connected at the joints of $\cMAT$. Each spline curve $C_j$ parameterizes both the position and the radius of medial points. Hence, any point on $C_j$ gives a medial point on $\cM$.  Similarly, each $C_j$ is a maximal curve segment comprising regular medial points and it corresponds to a branch in $\cMAT$.  An end-point of $C_j$ is either an end-point of $\cM$, or a joint of $\cM$ at which three of more branches meet (Fig.~\ref{fig:branchjoint:b}).

%
The \emph{envelope} $\widehat C_j$ of $C_j$  is specified as the union of all medial circles on $C_j$, noted that $C_j$ is a continuous curve.
Let $\widehat \cM$ be the union of $\{\widehat C_j\}$. Then $\widehat \cM$ represents a shape approximation to $\cO$.
Compared with discrete medial axis representation in which a 2D shape is approximated using a union of medial circles sampled on $\cMAT$, the piecewise smooth medial axis transform provides a compact yet accurate shape representation.  While a set of densely sampled medial circles needs to be sampled on a branch of $\cMAT$ to well approximate the shape the branch represents (Fig.~\ref{fig:branchjoint:c}), it takes far fewer control points to define spline curves to accurately approximate the medial axis transform.


\section{Computational Framework}\label{sec:compframe}

In this section we shall present a full pipeline for automatically generating a piecewise smooth medial axis transform for an arbitrary 2D geometric shape $\cO$ with smooth boundary $\partial \cO$. To start, we assume that $\cO$ is sampled by a set of dense sample points $\{\bvec{p}_i \in \mR^2\}$ on $\partial \cO$, meeting the sampling conditions~\cite{amenta1999surface} to faithfully capture the topology and features of the boundary $\partial \cO$.

Our algorithm consists of two main steps: (1) {\em noise filtering} for generating a stable medial axis transform $\cM_s$, and (2) {\em geometric simplification} for computing a spline-based medial axis transform $\cM$ based on the branch structure of $\cM_s$.
Our goal is to obtain $\cM$, whose reconstructed shape $\widehat \cM$ well approximates the original shape $\cO$, and both the above steps are governed by a quantitative measure $\varepsilon(\cO, \widehat\cM)$ of the approximation error of $\widehat \cM$ to $\cO$.

While the differences of two shapes are commonly measured by a Hausdorff distance (by considering the shapes as point sets), we use instead the one-sided Hausdorff distance from the original boundary $\partial \cO$ to the reconstructed boundary $\partial \widehat \cM$ of $\cM$ as the error measure $\varepsilon(\cO, \widehat\cM)$, which is easier to compute.  As $\partial \cO$ is generally more complex than the simplified $\partial \widehat\cM$, the one-sided Hausdorff distance is usually a good approximation to the Hausdorff distance of the boundaries of the two shapes.

Our algorithm is outlined in Algorithm~\ref{alg:framework}. We shall present the details in subsequent subsections.

\renewcommand{\algorithmicrequire}{\textbf{Input:}}
\renewcommand{\algorithmicensure}{\textbf{Output:}}

\begin{algorithm}[htbp!]
\caption{\textbf{Spline approximation to the medial axis transform of a 2D shape}}
\label{alg:framework}
\begin{algorithmic}[1]
\REQUIRE A dense set of sample points $\{\bvec{p}_i\}$ representing the boundary $\partial \cO$ of a 2D shape $\cO$, an error threshold $\hat \varepsilon$.
\ENSURE A piecewise smooth medial axis transform $\cM$ such that the approximation error $\varepsilon(\cO, \widehat\cM) \leq \hat \varepsilon$.
\STATE Obtain an initial medial axis transform $\cM_0$ from $\{\bvec{p}_i\}$ using the Voronoi-based approach~\cite{attali1997computing}.
\STATE {\em Noise pruning} -- Compute a stable medial axis transform $\cM_s$ from $\cM_0$ guided by the one-sided Hausdorff distance filtering, with the control of $\hat \varepsilon$.
\STATE {\em Geometric simplification} -- Obtain an initial piecewise smooth medial axis transform $\cM$ by spline fitting to $\cM_s$.  Optimize $\cM$ iteratively by minimizing an objective function characterizing the approximation error $\varepsilon(\cO, \widehat\cM)$.
\STATE Output the obtained $\cM$.
\end{algorithmic}
\end{algorithm}

\subsection{Noise filtering} \label{sec:noisefiltering} 
In this step, we aim at generating a stable medial axis transform $\cM_s$ for an input 2D shape $\cO$.
Furthermore, we ensure that the approximation error of the reconstructed shape $\widehat \cM_s$ to $\cO$, denoted as $\varepsilon(\cO, \widehat\cM_s)$, is no larger than a user specified error threshold  $\hat \varepsilon$. In our framework, $\varepsilon(\cO, \widehat\cM_s)$ is measured by the one-sided Hausdorff distance from $\partial \cO$ to the boundary $\partial \widehat \cM_s$ of $\widehat\cM_s$.
By noise pruning, we shall obtain a stable medial axis transform $\cM_s$ that will be used as an input in the next step for computing $\cM$, a smooth spline representation of the medial axis transform.

Given a dense set of points $\{\bvec{p}_i\}$ sampled on the boundary $\partial \cO$ of $\cO$,
we first apply the Voronoi diagram based method~\cite{attali1997computing}
to obtain an initial medial axis transform $\cM_0$ whose medial axis comprises the internal vertices and edges of the Voronoi diagram of $\{\bvec{p}_i\}$. The filtering then proceeds by considering the approximation error induced by the removal of medial points on the medial axis transform, as measured by the one-sided Hausdorff distance from $\partial \cO$ to $\partial \widehat \cM_s$.

The one-sided Hausdorff distance from $\partial \cO$ to $\partial \widehat\cM_s$ is defined as:
\begin{equation}
\varepsilon(\cO, \widehat\cM_s) = \max _i \{d(\bvec{p}_i, \partial \widehat \cM_s)\},
\end{equation}
where $d(\bvec{p}_i, \partial \widehat \cM_s)$ is the Euclidean distance from boundary point $\bvec{p}_i$ to $\partial \widehat \cM_s$.
The reconstructed shape $\widehat \cM_s$ is represented as the union of all medial circles in $\cM_s$. Since the medial axis of $\cM_s$ is a subset of the Voronoi diagram of $\{\bvec{p}_i\}$, $\bvec{p}_i$ always lies on or outside medial circle $(\bvec{u}_j,r_j)$.
Hence, $d(\bvec{p}_i, \partial \widehat \cM_s)$ equals the distance from $\bvec{p}_i$ to its nearest medial circle $(\bvec{u}_j,r_j)$:
\begin{equation}
d(\bvec{p}_i, \partial \widehat \cM_s)\ = \min_j \{ d(\bvec{p}_i,\bvec{u}_j)- r_j \}.
\end{equation}

\subsubsection{Filtering strategy}
To ensure that $\cM_s$ has the same homotopy as the initial medial axis $\cM_0$, only the end-points of the medial axis transform might be pruned. The initial $\cM_0$ gives the exact reconstruction of the shape with respect to the sample points
$\{\bvec{p}_i\}$ and hence $\varepsilon(\cO, \widehat \cM_0) = 0$.
At each iteration of the filtering process,
we check the error induced by the removal of a medial end-point $\bvec{v}_j$,
which equals $\delta_j = \varepsilon(\cO, \widehat \cM_s')$,
where $\cM_s' = \cM_s \setminus \{\bvec{v}_j\}$.
If the induced error $\delta_j \geq \hat \varepsilon$, $\bvec{v}_j$ will be treated as a feature of the medial axis and kept in $\cM_s$. Otherwise, we treat $\bvec{v}_j$ as noise and prune it from $\cM_s$. The process is repeated until all end-points have been checked.
Algorithm~\ref{alg:topological} describes the steps of our noise pruning algorithm.

\begin{algorithm}
\caption{\textbf{Noise pruning algorithm}}
\label{alg:topological}
\begin{algorithmic}[1]
\STATE Construct $\cM_0$ from the Voronoi diagram of $\{\bvec{p}_i\}$.
\STATE Set $\cM_s = \cM_0$, store all end-points in a queue $Q$.
\WHILE {$Q$ is not empty}
        \STATE Pop an end-point $\bvec{v}_j$ from $Q$.
        \STATE If $\delta_j < \hat \varepsilon$, remove $\bvec{v}_j$, as well as its neighboring segment from $\cM_s$ and push the neighbor point of $\bvec{v}_j$ into $Q$ if it becomes an end-point.
\ENDWHILE
\STATE \textbf{return} the stable medial axis transform $\cM_s$.
\end{algorithmic}
\end{algorithm}

The result of the above pruning procedure is in general not unique, since it might depend on the order of pruning. Nevertheless any such result will meet the specified error tolerance $\hat \varepsilon$, and can be used at the next stage of geometric simplification as an input to further compute a compact piecewise smooth medial axis transform.

\subsection{Geometric simplification} \label{sec:geosimp}
After noise pruning, a stable medial axis transform is obtained. However, the pruning step removes only noisy medial points and leaves numerous discrete medial points on the stable medial branches. While those medial points are critical to a faithful representation for union of circles, in the envelope representation which we adopt, most of them contribute quite little to the shape and are thus nearly redundant. To get a concise medial axis transform, we should further reduce the number of medial points by utilizing a piecewise continuous representation. Such medial point decimation is driven by a user-specified approximation error threshold $\hat \varepsilon$ to ensure the approximation error is under control.

First, we construct an undirected graph $\mc{G}$ to represent the medial points filtered by noise pruning algorithm. Each medial point corresponds to one vertex in $\mc{G}$. Two vertices in $\mc{G}$ are connected by an edge if their corresponding medial points are neighbors in the medial axis transform.
We then obtain chains from the graph by grouping all edges which are connected without passing through a joint. Since each chain $\mc{H}_j$ is single connected, we fit a smooth curve $C_j$ for it. By considering the smoothness, the ease of implementation and the representation ability, we choose the cubic B-spline to represent the smooth curve for each chain.



\subsubsection{Piecewise cubic B-spline medial axis transformation initialization}
We first sample dense points from the medial axis transform and find an initial cubic B-spline medial axis transform with open-uniform knot vectors to fit these medial points~\cite{eberly2008}. The reason that we choose this fitting algorithm is, it could generate good fitting curves only based on the number of control points; no initial curve is required. To ensure that the B-spline curve passes through the end-points, knots of multiplicity four are used at each end-point.


The fitting error of the $C_j$ curve is measured by the maximal Euclidean distance from medial points on chain $\mc{H}_j$ to their projections on the fitting curve $C_j$. Let $f_e$ be the maximal fitting error over all spline curves.
Ji{\v{r}}{\'\i} Kosinka and Bert J{\"u}ttler~\cite{kosinka2006g1} proved that,
the boundary approximation error $\varepsilon(\cO, \widehat\cM)$, measured by the one-sided Hausdorff distance from the original boundary $\partial \cO$ to the reconstructed boundary $\partial \widehat \cM$ from $\cM$, is upper bounded by $\sqrt{2} \cdot f_e$.
Hence, we mark a spline curve $C_j$ as \emph{reliable}, if the fitting error of $C_j$ does not exceed $\hat \varepsilon/\sqrt 2$.
For every branch $\mc{H}_j$ of $\cM_s$, our fitting strategy is to search a \emph{reliable} cubic B-spline curve with the minimal number of control points.

\subsubsection{Medial axis transform optimization}
Then, the distance between $\partial \cO$ and $\partial \widehat \cM$ is minimized by optimizing the coordinates of B-spline control points.
We'll define an energy function $E(\bvec X)$ to formulate this problem.
In the following presentation, the variables, $\bvec{X} = \{\bvec{x}_i\}$ are control points of $\{C_j\}$. Each $\bvec{x}_i = (\bvec{u}_i, r_i) \in \mR^3$, where $\bvec{u}_i$ and $r_i$ denote the position and the radius of a control point.

The distance between the two curves is approximated by the sum of the squared distances.
\begin{equation}\label{eq:2dmatenergy}
E(\bvec X) = \sum _{i = 1} ^n {d^2(\bvec{p}_i,\partial \widehat \cM)},
\end{equation}
where $d^2(\bvec{p}_i,\partial \widehat \cM)$ is the squared Euclidean distance from $\bvec{p}_i$ to $\partial \widehat \cM$.
Since $\widehat \cM$ is the union of all $\widehat C_j$, $d^2(\bvec{p}_i,\partial \widehat \cM)$ could be computed by the
squared distance from $\bvec{p}_i$ to the boundary of its closest envelope $\widehat C_j$,
\begin{equation}
d^2(\bvec{p}_i,\partial \widehat \cM) = \min \limits_j d^2(\bvec{p}_i, \partial \widehat C_j),
\end{equation}
where $\partial \widehat C_j$ represents the boundary of an envelope $\widehat C_j$.

As $\widehat C_j$ is the union of all medial circles it represents, a bottom-up way to compute $d^2(\bvec{p}_i,  \partial \widehat C_j)$ is designed in our algorithm.
By evenly and densely sampling in the spline parameter space, a set of piecewise linear segments are obtained.
Let $S$ be a segment sampled on $C_j$.
Then its reconstruction $\widehat S$ is the envelope of its two medial circles, with a boundary $\partial \widehat S$ (see Fig.~\ref{fig:envelope}).
If the sampling density in spline parameter space is high enough, $\widehat C_j$ can be approximated accurately as a union of all such envelopes $\widehat S$.
We consider the footpoint of $\bvec{p}_i$ on one envelope $\widehat S$ as invalid, if the footpoint lies in the interior of another envelope.
In that case, $d^2(\bvec{p}_i, \partial \widehat C_j)$ is the minimal valid $d^2(\bvec{p}_i, \partial \widehat S)$. We will provide the details for computing $d^2(\bvec{p}_i, \partial \widehat S)$ in~\ref{sec:app:errorterm}.

By minimizing $E$, an optimized $\cM$ is obtained. Intuitively, more control points in $\cM$ will lead to a smaller $E$ and a smaller $\varepsilon(\cO, \widehat \cM)$.
As we know, the energy is defined in the $L^2$ space, while the approximation error $\varepsilon(\cO, \widehat \cM)$ is defined in the $L^\infty$ space. It is possible that even if $E$ converges to a minimal value, $\varepsilon(\cO, \widehat \cM)$ is still larger than $\hat \varepsilon$. If that happens, our strategy is to insert new control points in $\cM$ until $\varepsilon(\cO, \widehat \cM)$ is no greater than $\hat \varepsilon$.

In our algorithm, $\varepsilon(\cO, \widehat \cM)$ is compared with $\hat \varepsilon$, at the end of the optimization process. If $\varepsilon(\cO, \widehat \cM) > \hat \varepsilon $, our algorithm will pick up the boundary point $\bvec{p}_k$ with the largest distance to $\partial \widehat \cM$, and insert a new control point to the spline curve whose envelope $\bvec{p}_k$'s footpoint lies on. Another round of optimization will be executed, and these steps are repeated until a medial axis transform with a satisfactory approximation error is achieved. The steps are described in Algorithm~\ref{alg:geosimp}.

\begin{algorithm}
\caption{\textbf{Geometric simplification algorithm}}
\label{alg:geosimp}
\begin{algorithmic}[1]
\STATE Obtain an initial $\cM$ by fitting the medial points in $\cM_s$ with a set of cubic B-spline curves $\{C_j\}$.
\STATE Optimize $\cM$ by minimizing the energy function $E$.
\WHILE {$\varepsilon(\cO, \widehat \cM) > \hat \varepsilon$}
        \STATE Insert a new control point in $\cM$, and minimize $E$ again.
\ENDWHILE
\STATE \textbf{return} a medial axis transform $\cM$ with $\varepsilon(\cO, \widehat \cM) \leq \hat \varepsilon$.
\end{algorithmic}
\end{algorithm}

When inserting a new control point to a spline curve, the user is not required to specify the location of the new control point, but the spline curve fitting algorithm will compute and adjust the location based on the number of control points, which is similar to the spline curve initialization.


\section{Implementation Details} \label{sec:implementation}
In this section, some implementation details will be discussed.\\

\noindent{\em Initial sampling.}
To get a good approximation of the medial axis transform, the input sample points $\{\bvec{p}_i\}$ should capture the boundary topology faithfully, as well as boundary features.
The sampling condition we adopt is the local feature size condition discussed in~\cite{amenta1999surface}. The local feature size at a boundary point $\bvec{p}_i$ is the Euclidean distance from $\bvec{p}_i$ to its nearest point on the medial axis. In our case, if the boundary is $r$-sampled for $r \leq 0.25$, we consider the set $\{\bvec{p}_i\}$ as a valid input.
\\

\noindent{\em Optimization.}
We apply the L-BFGS method~\cite{liu1989limited}, an iterative quasi-Newton method, to minimize the energy function $E(\bvec{X})$. The L-BFGS method takes the control points $\bvec{X}$ in $\cM$ as the input, evaluates the objective function $E(\bvec{X})$ and its gradient $\nabla E(\bvec{X})$ in the optimization. Since no Hessian is involved in the computation, the L-BFGS method is quite efficient in finding the optimized medial axis transform.\\

\noindent{\em Global check and local check.}
There're overlaps among envelopes. A footpoint is only valid if it is not contained in any other envelope.
To determine the validity of a footpoint, a global check is required to traverse all envelopes, which is quite time-consuming as a repeated step in each iteration. We find that generally the footpoint will move only marginally after each iteration. The observation inspires us to adopt a hybrid check strategy, which uses the local check and calibrates the footpoint projection with a global check after a certain number of iterations.\\

\noindent{\em Medial axis transform validation.}
The case where one medial circle is completely contained by another medial circle is not allowed in a valid medial axis transform. Let $\bvec{v}_1=(x_1,y_1,r_1), \bvec{v}_2=(x_2,y_2,r_2)$ be any two different medial points in a medial axis transform, and $\Delta x = x_1 - x_2, \; \Delta y = y_1 - y_2, \; \Delta r = |r_1 - r_2|$. In a valid medial axis transform, the slope of the two points, defined as
\begin{equation}\label{eq:slope}
\tan \alpha = \frac{\Delta r}{\sqrt{\Delta x ^2 + \Delta y ^2}},
\end{equation}
should satisfy $\|\tan \alpha \| \leq 1$. Although we do not incorporate this validity constraint in our optimization for the sake of simplicity, we have checked our computed examples and found no violation. However, to guarantee the validity explicitly, it is an interesting future problem to implement efficient constrained optimization to include this validity condition.


\section{Experiments and Discussions} \label{sec:experiments}
In this section, we choose some data sets and show how our algorithm performs on these shapes.
All experiments are conducted on a Xeon 3.33 GHz PC with 12 GB RAM. We implement our system in C\texttt{++} and adopt CGAL~\cite{cgal} to compute the Delaunay triangulation and generate the Voronoi diagram of sampling points on the shape boundary. The $L^\infty$ approximation error $\varepsilon$ and error threshold $\hat \varepsilon$ are normalized by scaling the diagonal of the bounding box of the input shape to $1$.

In all figures presented in this paper, the blue regions are the reconstructed shapes from medial axis transforms and the black contours are boundaries of the input shapes.
Black dots in spline medial axis transforms represent the positions of the B-spline control points. In other discrete medial axis transforms, black dots are the centers of medial circles.

\subsection{Workflow}
\begin{figure*}[ht]
\centering
    \begin{subfigure}[t]{0.24\textwidth}
    \centering
    \includegraphics[width=\textwidth]{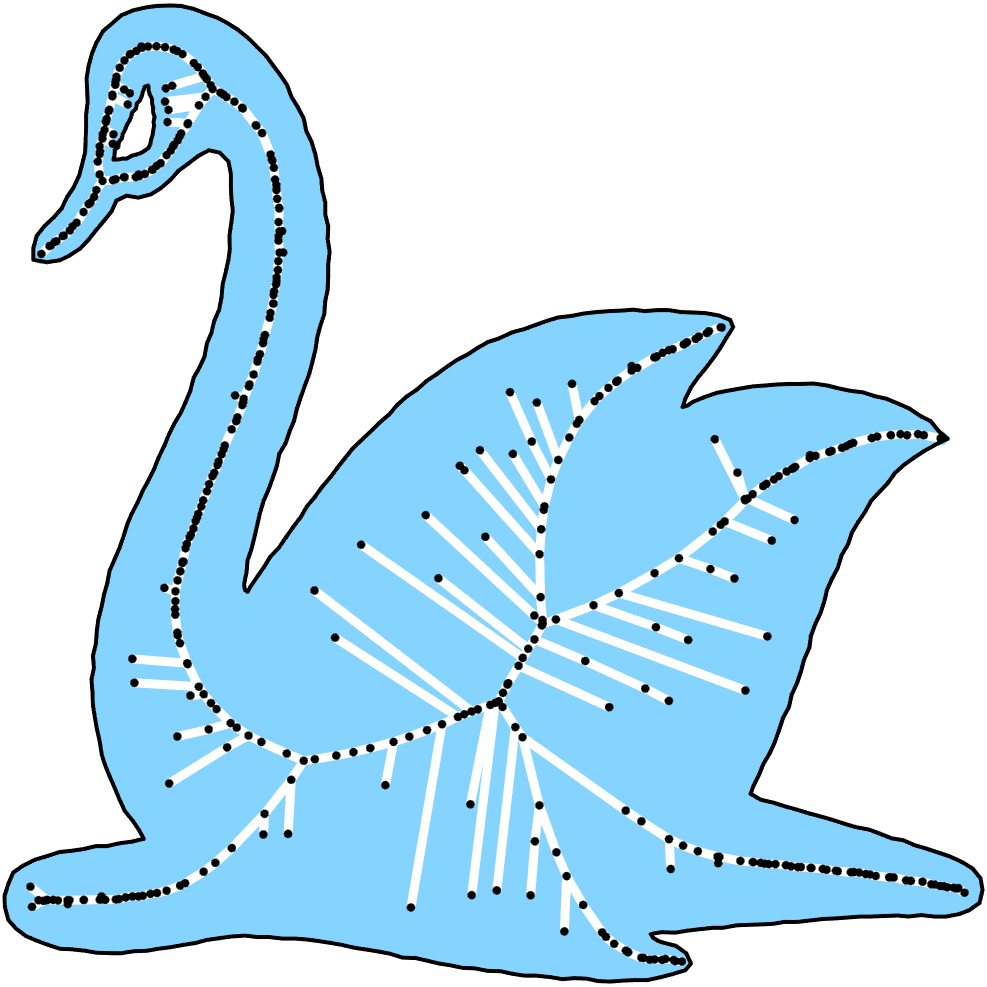}\\
    \caption{}
    \label{fig:swancomp:raw}
    \end{subfigure}
    \hfill
    \begin{subfigure}[t]{0.24\textwidth}
    \centering
    \includegraphics[width=\textwidth]{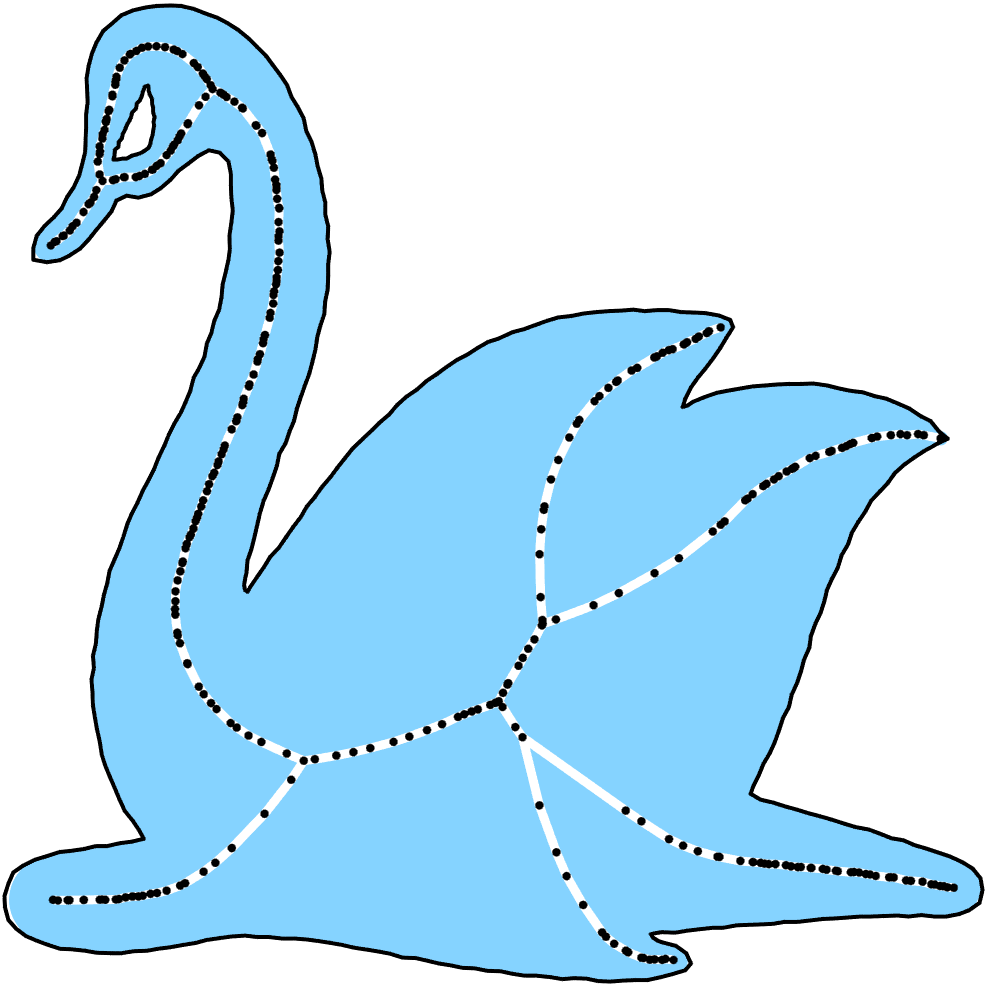}
    \caption{}
    \label{fig:swancomp:dis}
    \end{subfigure}
    \hfill
    \begin{subfigure}[t]{0.24\textwidth}
    \centering
    \includegraphics[width=\textwidth]{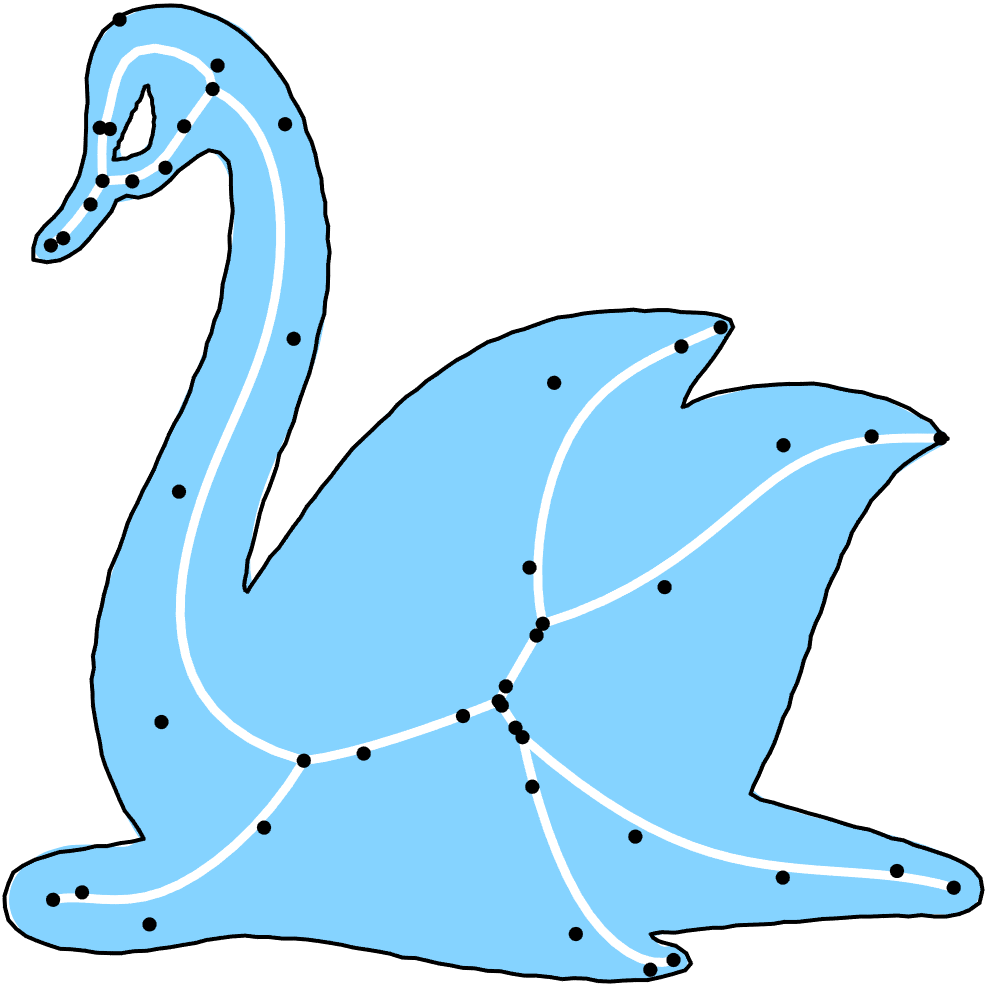}
    \caption{}
    \label{fig:swancomp:init}
    \end{subfigure}
    \hfill
    \begin{subfigure}[t]{0.24\textwidth}
    \centering
    \includegraphics[width=\textwidth]{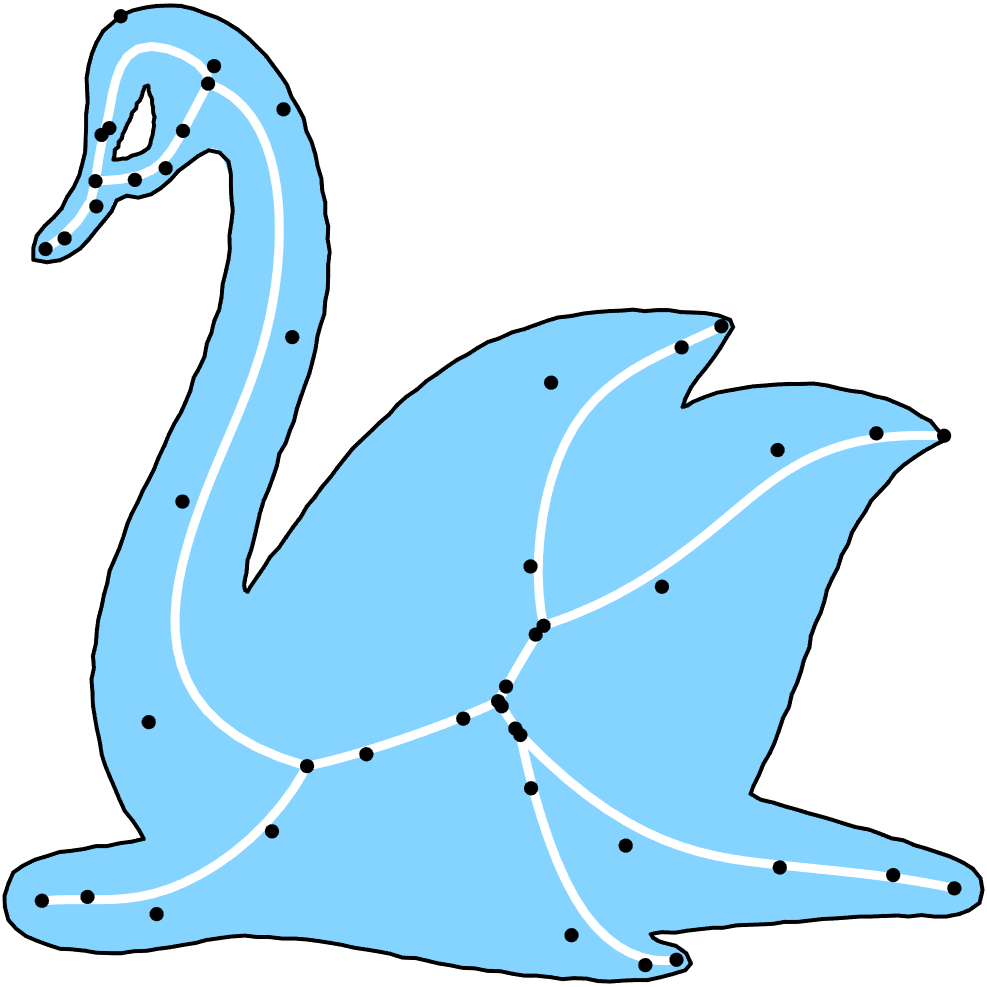}
    \caption{}
    \label{fig:swancomp:cb}
    \end{subfigure}
\caption{\small The workflow of our algorithm. (a) The initial medial axis transform, $|V|=445$. (b) A stable medial axis transform, $|V|=357$, $\varepsilon=0.40\%$. (c) A piecewise cubic B-spline medial axis transform before optimization. (d) A piecewise cubic B-spline medial axis transform after optimization, $|V|=46$, $\varepsilon=0.27\%$. In (a) and (b), $|V|$ is the number of medial points, while in (c) and (d), $|V|$ is the number of control points in cubic B-spline curves. The 2D positions of those points are rendered as black dots.}
\label{fig:swancomp}
\end{figure*}

The workflow of our algorithm, from the initial medial axis transform to its piecewise smooth medial axis transform, is shown in Fig.~\ref{fig:swancomp}.
%
The initial medial axis transform is computed based on the Voronoi diagram of boundary points (Fig.~\ref{fig:swancomp:raw}).
Filtering the noise in Fig.~\ref{fig:swancomp:raw} with an error threshold $\hat \varepsilon = 0.40\%$, we obtain a clean but discrete sample points of the medial axis transform (Fig.~\ref{fig:swancomp:dis}), with 357 medial points and approximation error $\varepsilon = 0.40\%$.
In Fig.~\ref{fig:swancomp:init}, 12 cubic B-spline curves with 46 control points are used to fit medial points in the discrete medial axis transform, and an initial piecewise smooth medial axis transform is ready.
After optimization, a piecewise cubic B-spline medial axis transform, whose approximation error $\varepsilon = 0.27\%$, is shown in Fig.~\ref{fig:swancomp:cb}.
This example shows that the spline representation computed by our algorithm not only provides a smooth and compact representations for medial axis transforms, but can also improve the approximation accuracy.

\subsection{Experiments on noise pruning}

\begin{figure*}[htbp]
\begin{minipage}[b]{0.37\linewidth}
\centering
    \begin{subfigure}[t]{0.48\textwidth}
    \centering
    \includegraphics[scale=0.19]{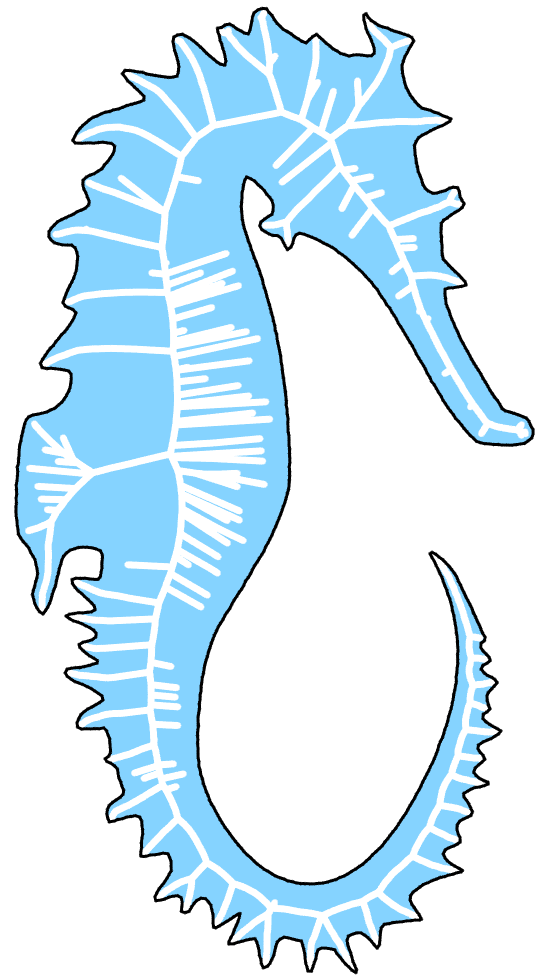}
    \caption{}
    \label{fig:seahorse:raw}
    \end{subfigure}
    \hfill
    \begin{subfigure}[t]{0.48\textwidth}
    \centering
	\includegraphics[scale=0.19]{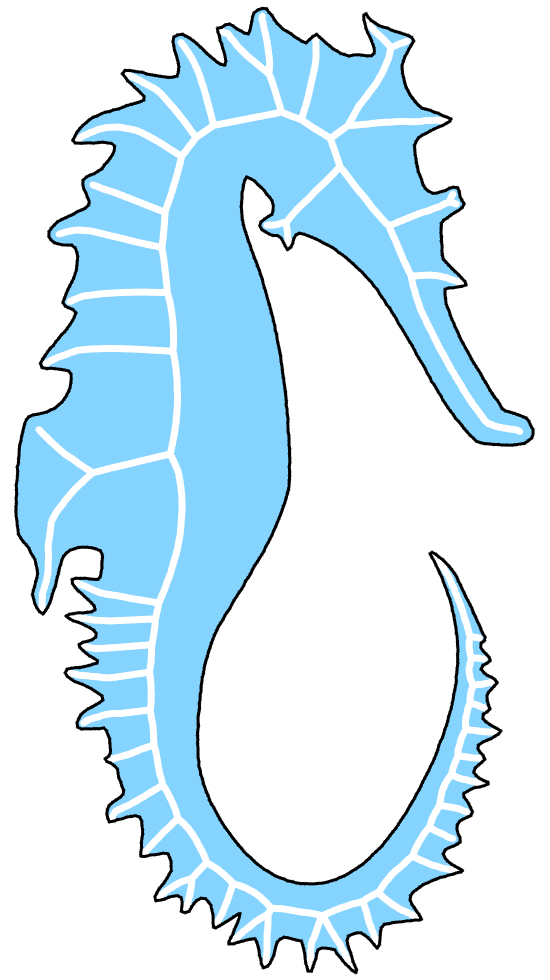}
    \caption{}
    \label{fig:seahorse:dis}
    \end{subfigure}
\caption{\small (a) The initial medial axis transform of a seahorse. (b) A stable medial axis transform via noise pruning with $\hat \varepsilon = 0.10\%$, and its approximation error $\varepsilon = 0.09\%$. } \label{fig:seahorse}
\end{minipage}
\hfill
\begin{minipage}[b]{0.60\linewidth}
\centering
    \begin{subfigure}[t]{0.48\textwidth}
    \centering
    \includegraphics[width=\textwidth]{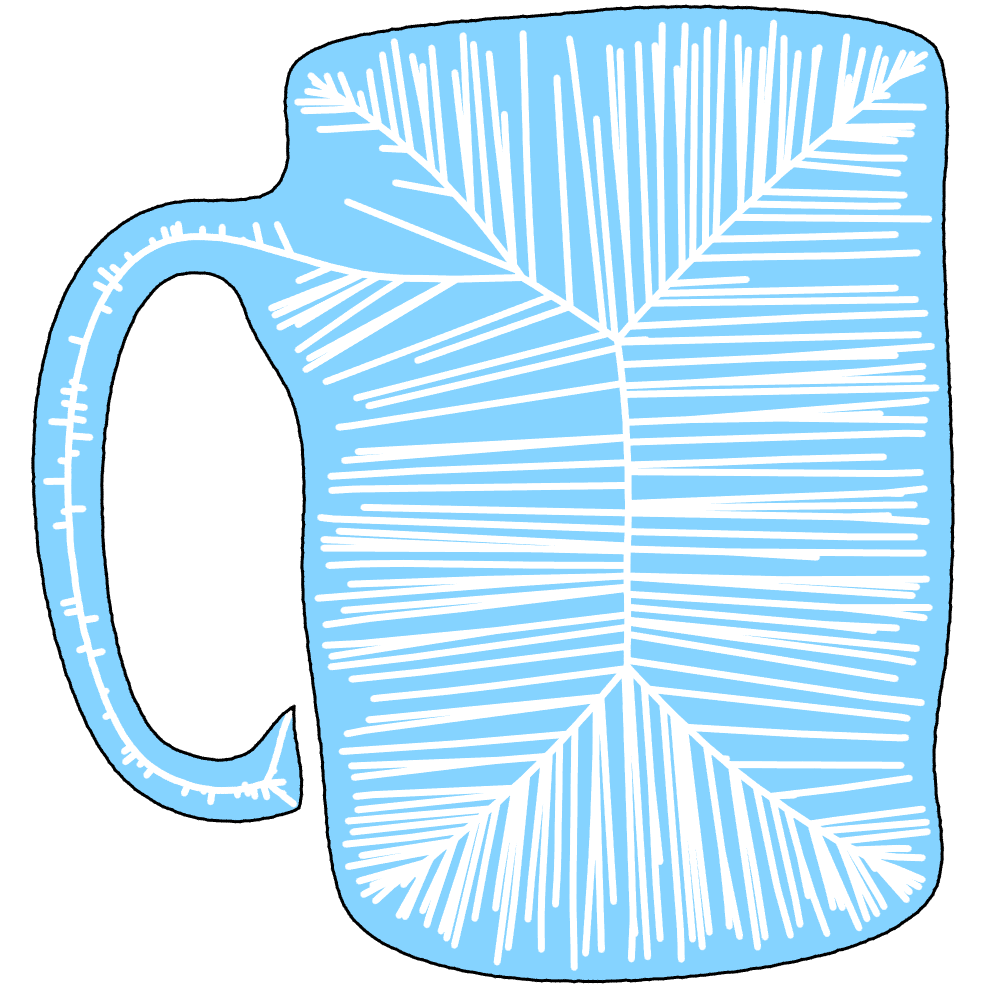}
    \caption{}
    \label{fig:mug:raw}
    \end{subfigure}
    \hfill
    \begin{subfigure}[t]{0.48\textwidth}
    \centering
    \includegraphics[width=\textwidth]{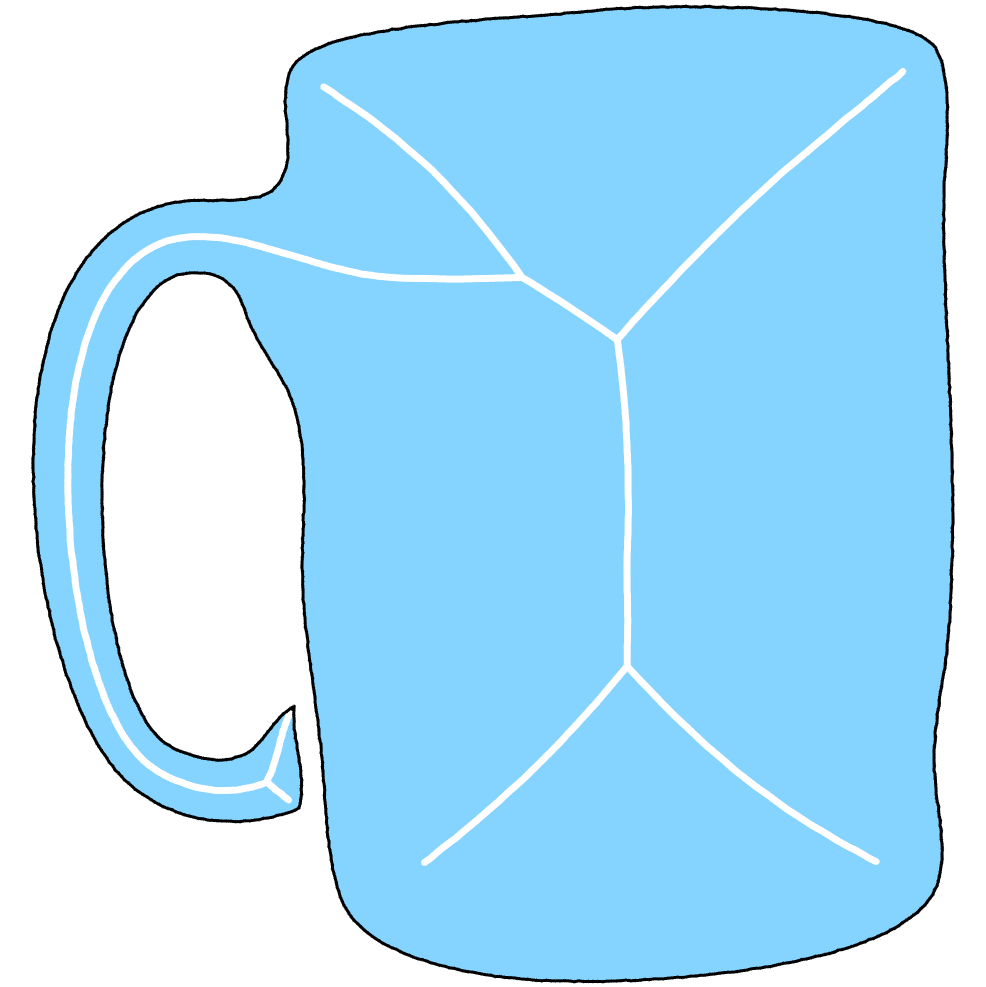}
    \caption{}
    \label{fig:mug:dis}
    \end{subfigure}
\caption{\small Perform noise pruning algorithm on a mug with perturbed boundary. (a) The initial medial axis transform which contains numerous noisy branches. (b) A resulting stable medial axis transform via noise pruning with $\hat \varepsilon = 0.07\%$, and its approximation error $\varepsilon = 0.07\%$. }
\label{fig:mug}
\end{minipage}
\end{figure*}

Our experiments demonstrate that the noise filtering algorithm (Algorithm~\ref{alg:topological}) provides a reliable and stable medial axis transform with guaranteed approximation error. An examples is shown in Fig.~\ref{fig:seahorse}. In Fig.~\ref{fig:seahorse:dis}, the noisy branches in the initial medial axis have been filtered successfully.
Meanwhile, the approximation error of the stable medial axis transform, $0.09\%$, is less than the specified error threshold $0.10\%$.

When handling shapes with perturbations on the boundary, our pruning algorithm also performs well. Although the initial medial axis in Fig.~\ref{fig:mug:raw} contains numerous undesired branches, the resulting medial axis in Fig.~\ref{fig:mug:dis} is topologically clean and stable. More noise pruning results are displayed in Fig.~\ref{fig:performances}. This fact shows that our noise pruning algorithm is a convincing method to deal with noisy branches in the medial axis transform.

\begin{figure*}[htbp]
\centering
    \begin{subfigure}[t]{0.32\textwidth}
    \centering
    \includegraphics[width=\textwidth]{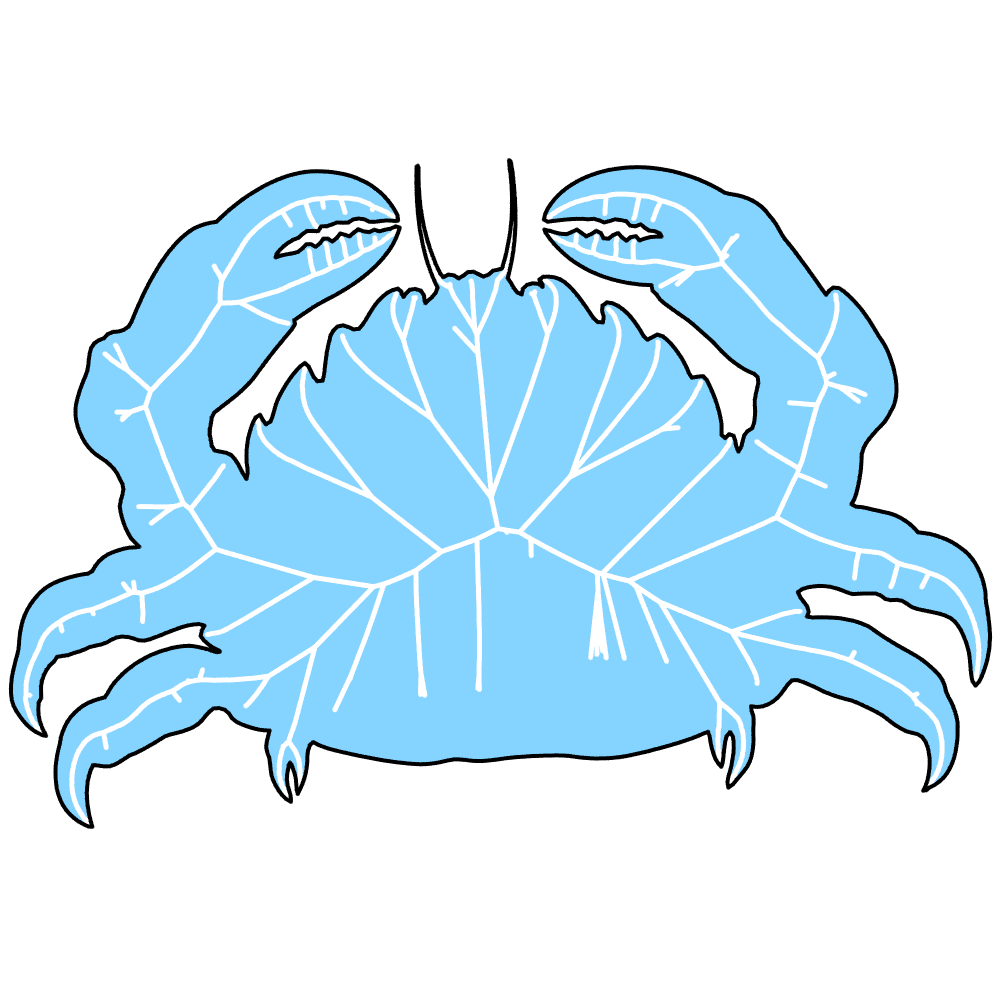}
    \caption{}
    \label{fig:crabs:raw}
    \end{subfigure}
    \hfill
    \begin{subfigure}[t]{0.32\textwidth}
    \centering
	\includegraphics[width=\textwidth]{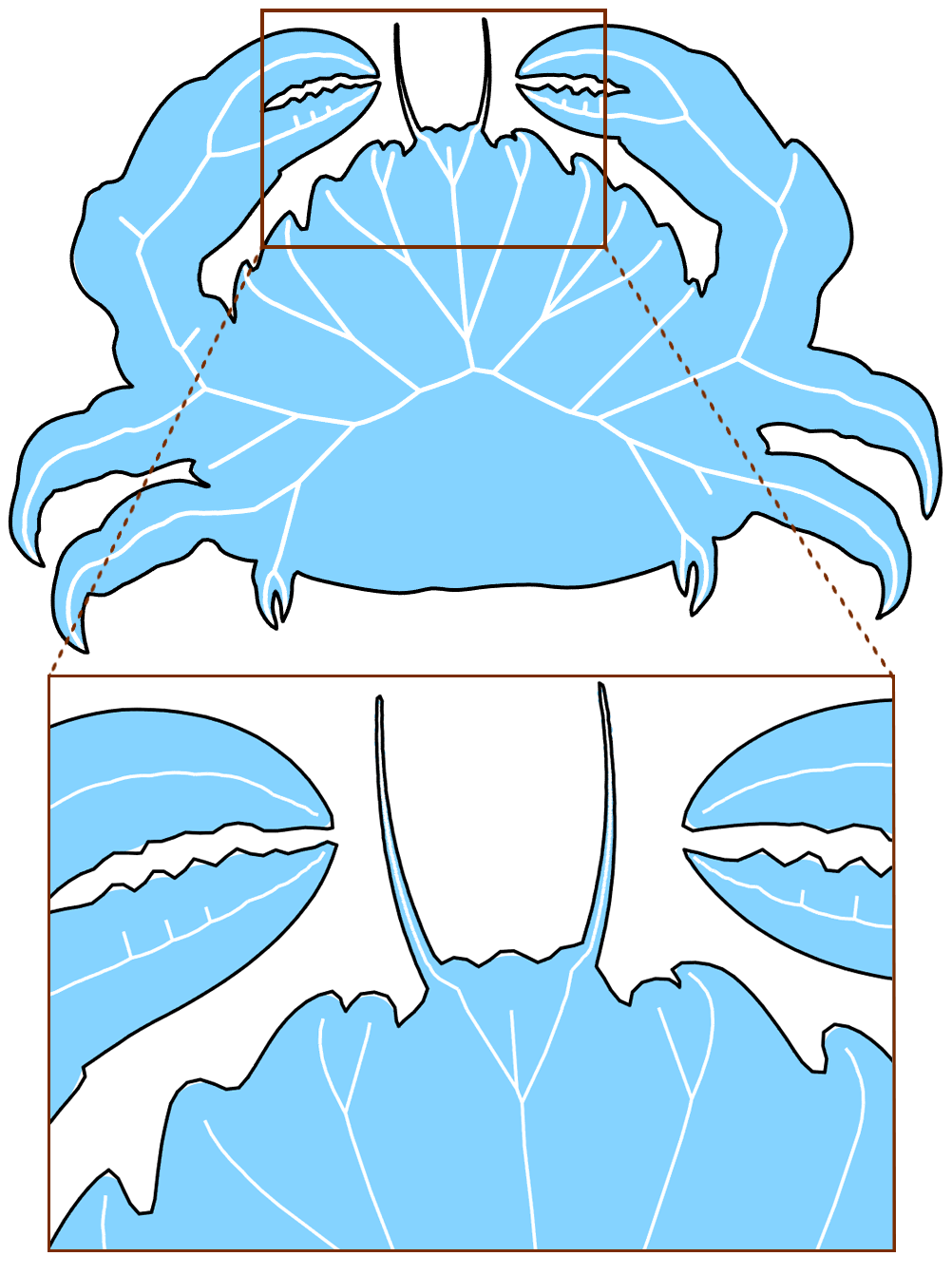}
    \caption{}
    \label{fig:crabs:small}
    \end{subfigure}
    \hfill
    \begin{subfigure}[t]{0.32\textwidth}
    \centering
	\includegraphics[width=\textwidth]{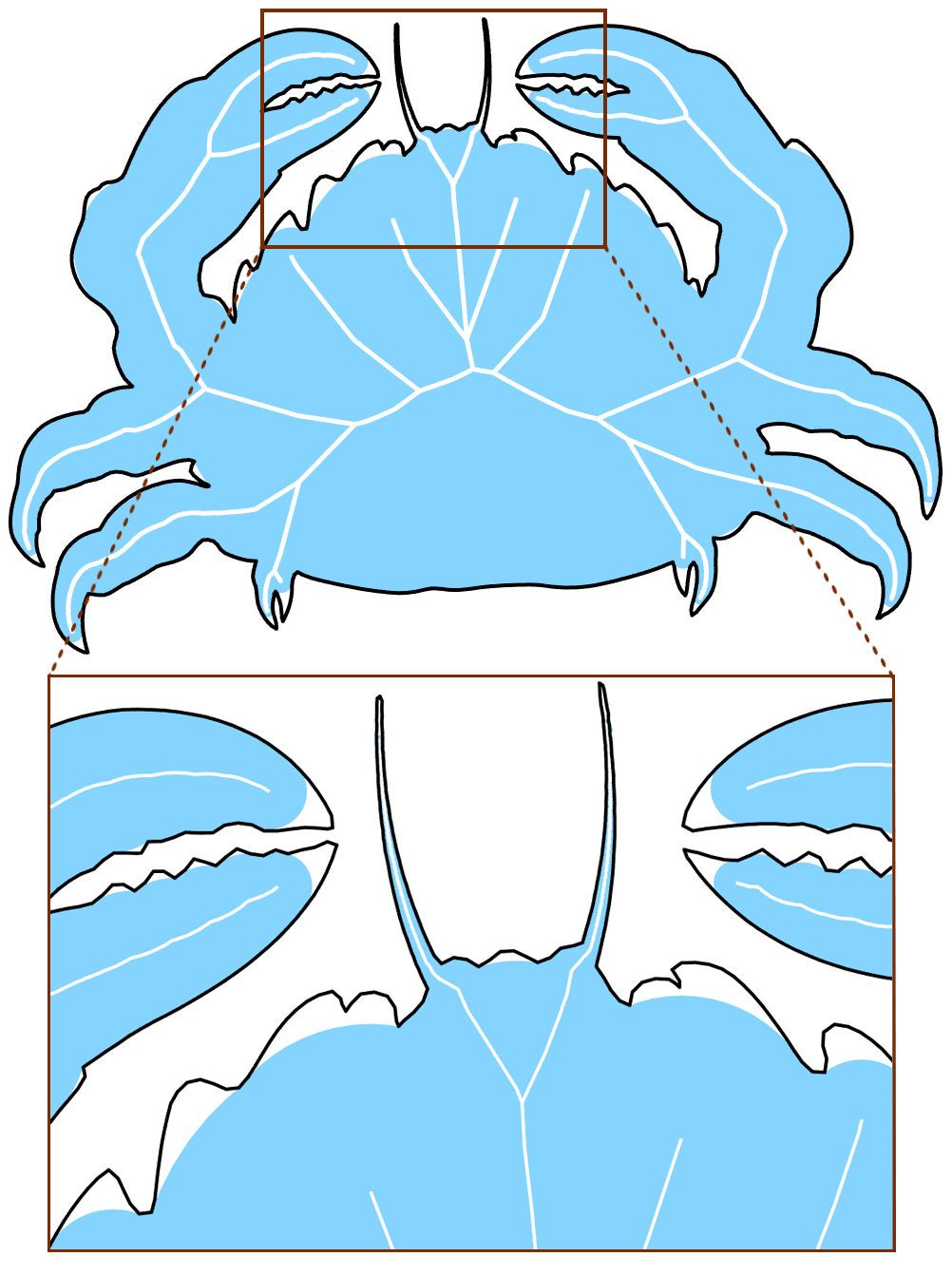}
    \caption{}
    \label{fig:crabs:large}
    \end{subfigure}
\caption{\small A comparison of different error thresholds in our noise pruning algorithm (Algorithm~\ref{alg:topological}). (a) The initial medial axis transform of a crab. (b) A stable medial axis transform via noise filtering with a smaller error threshold $\hat \varepsilon = 0.28\%$. (c) A stable medial axis transform via noise filtering with a larger error threshold $\hat \varepsilon = 1.40\%$. For illustration purpose, zoom-in views are displayed in the bottom row.}
\label{fig:crabs}
\end{figure*}

The error threshold can be fine-tuned to obtain different levels of details to be preserved.
If the user would like to preserve more details of the input shape, a smaller error threshold should be applied, as shown in Fig.~\ref{fig:crabs:small}. When a relatively large error threshold is applied, the main skeleton of the shape becomes clear, as shown in Fig.~\ref{fig:crabs:large}.


\subsubsection{Comparisons with other pruning strategies}

\begin{figure*}[htbp]
\centering
    \begin{subfigure}[t]{0.18\textwidth}
    \centering
    \includegraphics[width=\textwidth]{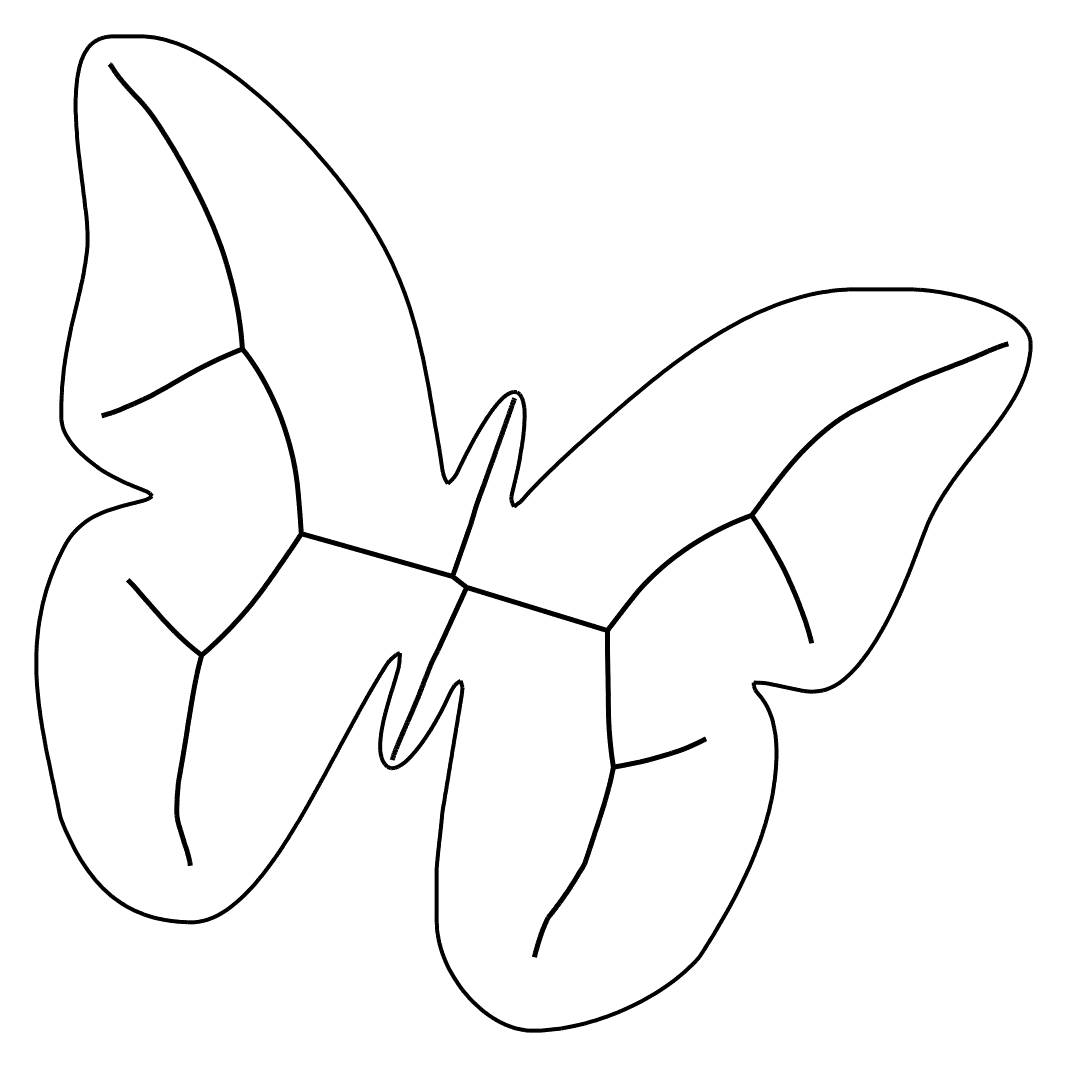}
    \caption{}
    \label{fig:butterflys:smooth}
    \end{subfigure}
    \hfill
    \begin{subfigure}[t]{0.18\textwidth}
    \centering
    \includegraphics[width=\textwidth]{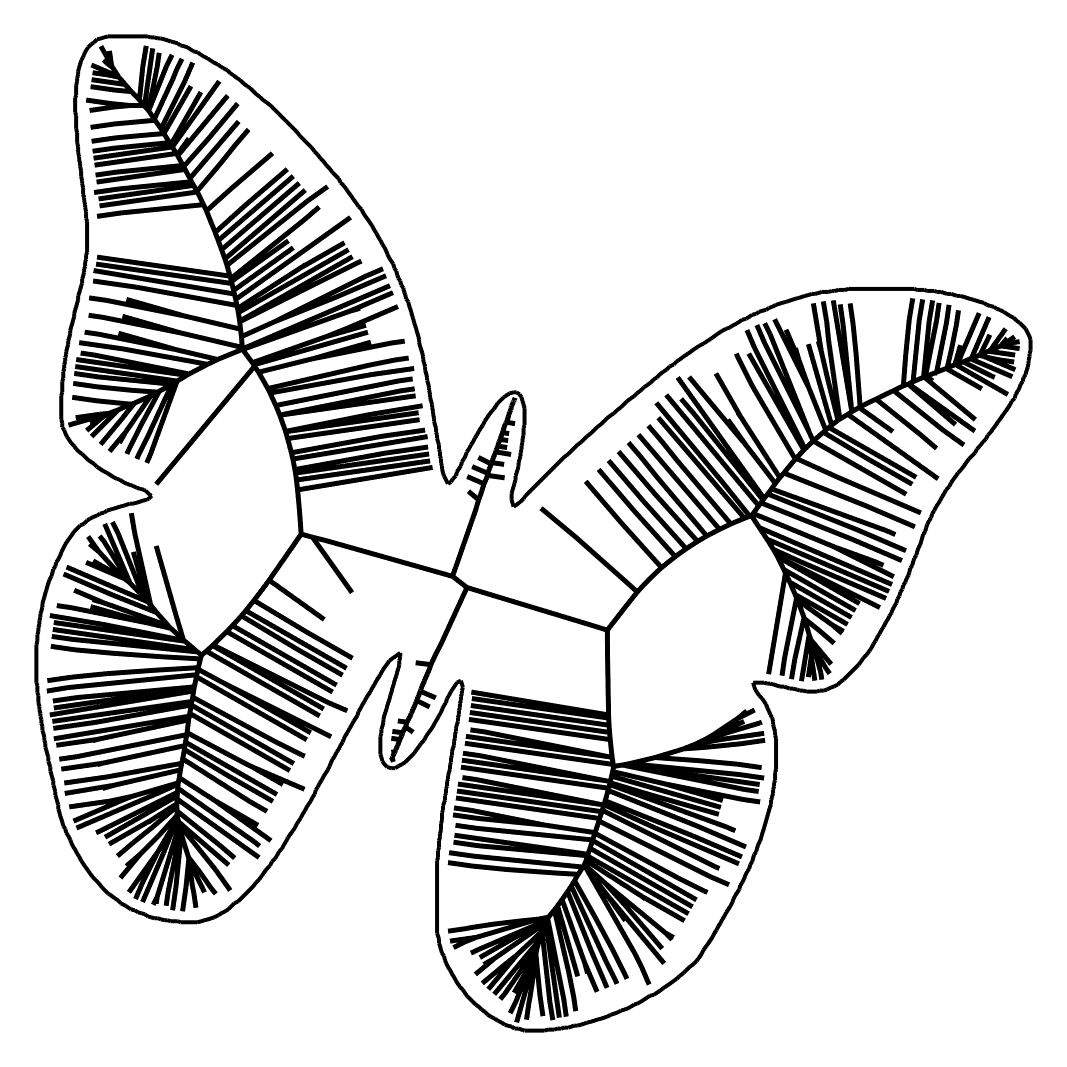}
    \caption{}
    \label{fig:butterflys:raw}
    \end{subfigure}
    \hfill
    \begin{subfigure}[t]{0.18\textwidth}
    \centering
	\includegraphics[width=\textwidth]{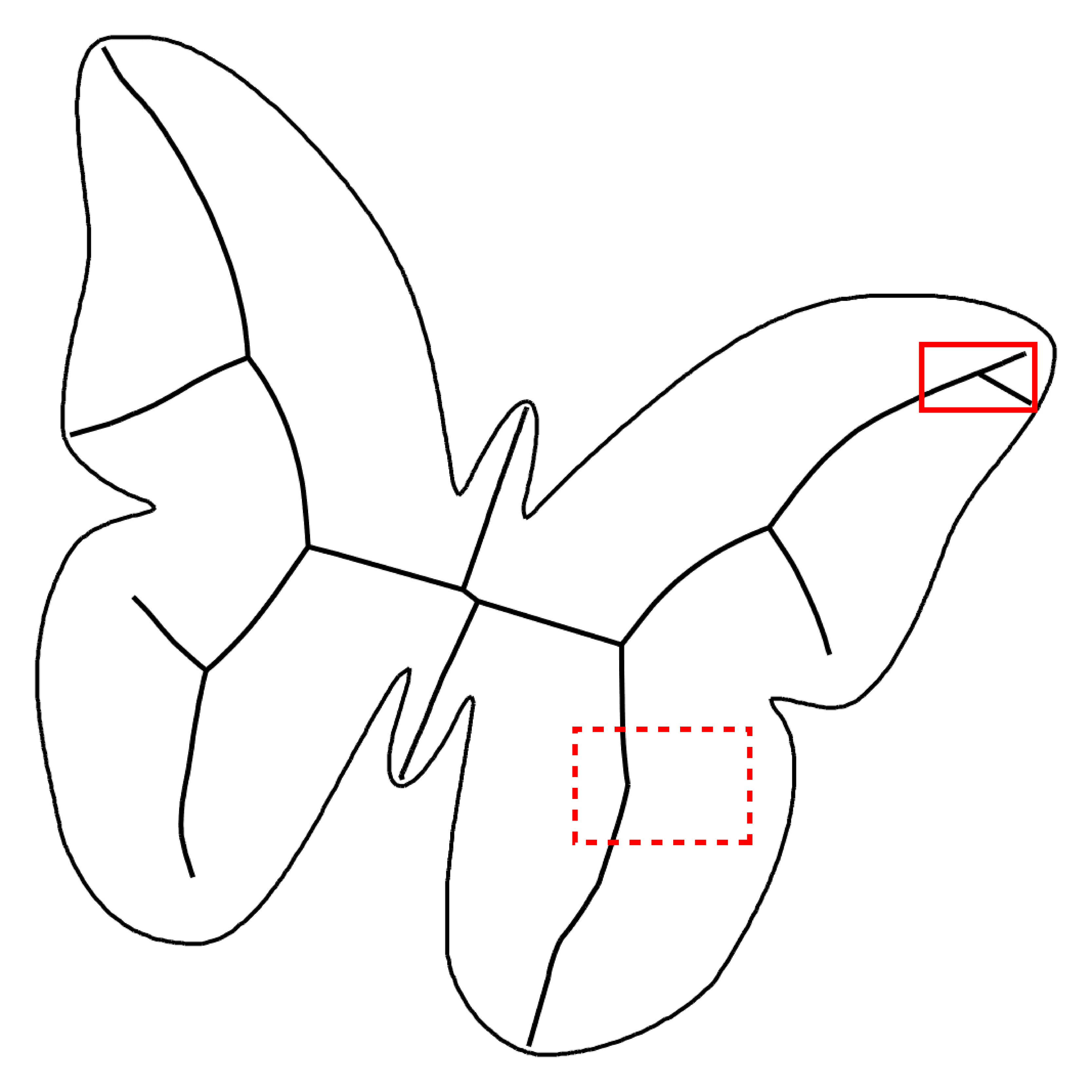}
    \caption{}
    \label{fig:butterflys:angle}
    \end{subfigure}
    \hfill
    \begin{subfigure}[t]{0.23\textwidth}
    \centering
	\includegraphics[width=\textwidth]{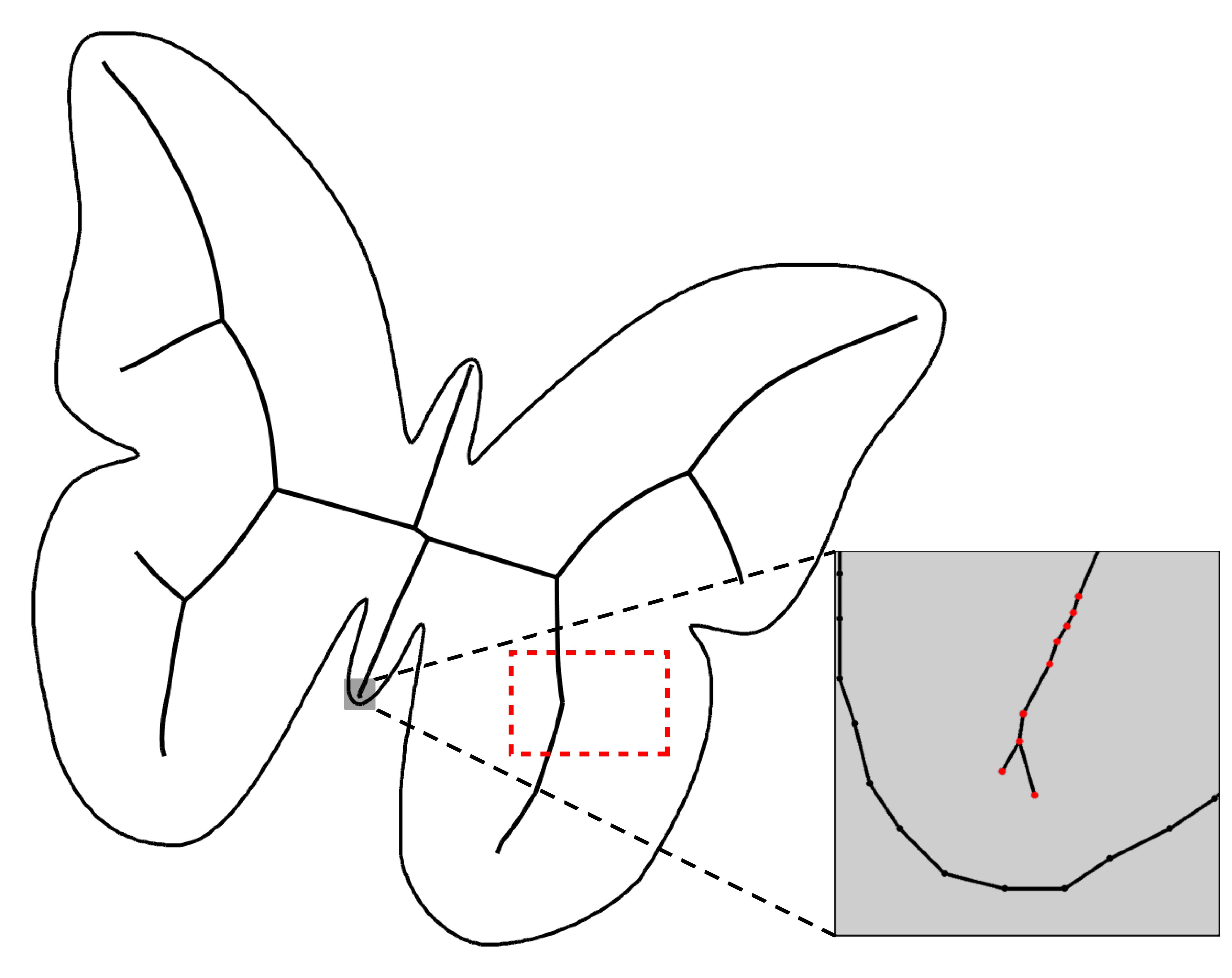}
    \caption{}
    \label{fig:butterflys:sat}
    \end{subfigure}
    \hfill
    \begin{subfigure}[t]{0.18\textwidth}
    \centering
	\includegraphics[width=\textwidth]{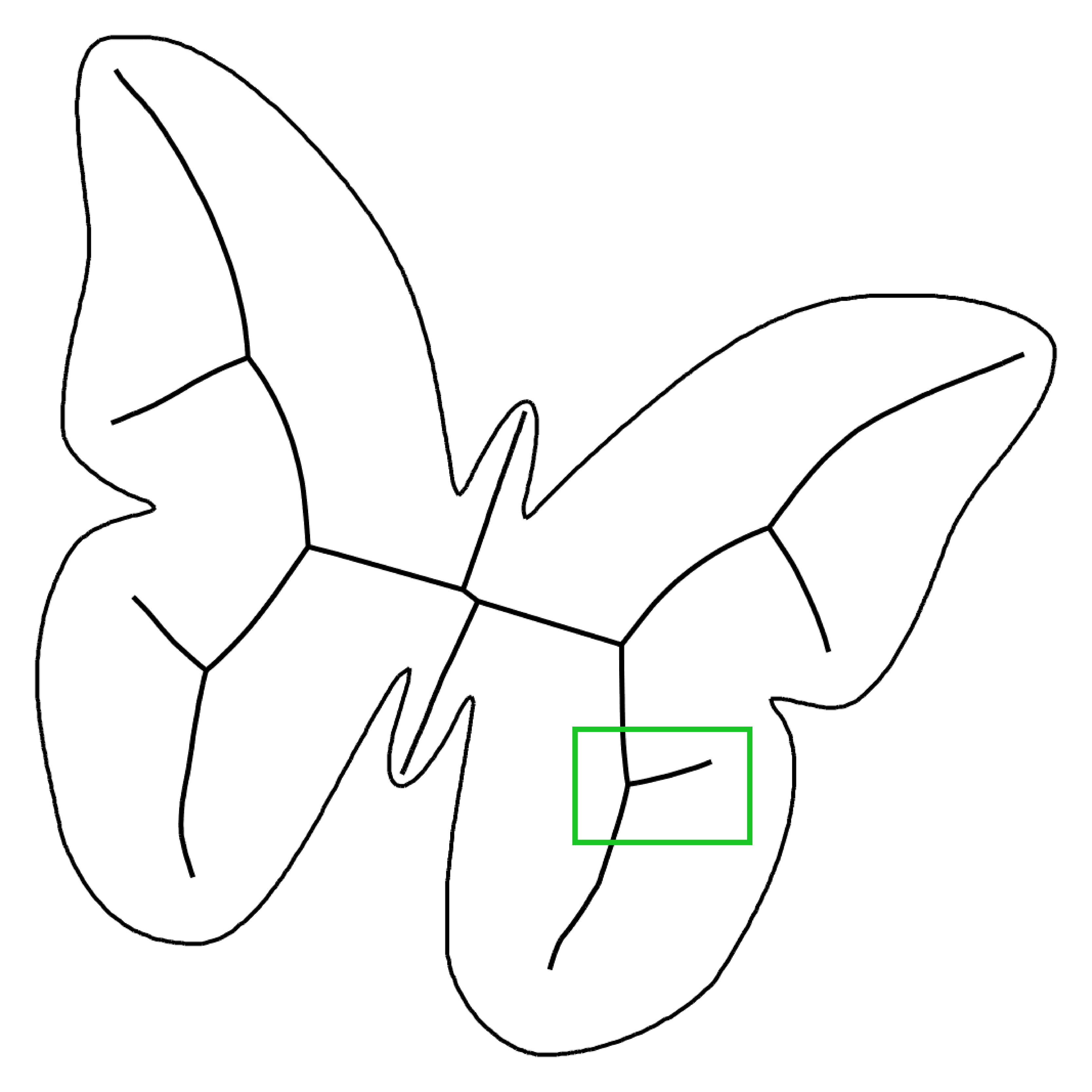}
    \caption{}
    \label{fig:butterflys:dis}
    \end{subfigure}
\caption{\small A comparison of medial axis pruning methods. (a) The ground truth. (b) The initial medial axis of a butterfly with noisy boundary. (c) The angle-based method~\cite{Sud:2005}, $\varepsilon = 0.34\%$. (d) The SAT method~\cite{giesen2009scale}, $\varepsilon = 0.34\%$. (e) Our algorithm, $\varepsilon = 0.10\%$.}
\label{fig:butterflys}
\end{figure*}

There are existing medial axis pruning methods to filter noise from the initial medial axis transform. We perform a comparison of our method with two typical methods -- the angle-based method and the scale axis transform (SAT) on the butterfly shape in Fig.~\ref{fig:butterflys}. The medial axis of a butterfly with a smooth boundary (Fig.~\ref{fig:butterflys:smooth}) is chosen as the ground truth. Fig.~\ref{fig:butterflys:raw} is generated after adding white noise on the smooth boundary in Fig.~\ref{fig:butterflys:smooth}, which results in a medial axis with many unstable branches. Fig.~\ref{fig:butterflys:angle} to~\ref{fig:butterflys:dis} show filtering results on Fig.~\ref{fig:butterflys:raw} by angle-based method~\cite{Sud:2005} with angle threshold 0.81, SAT~\cite{giesen2009scale} with scale parameter 1.12 and our method with $\hat \varepsilon = 0.10\%$, respectively.

It can be observed that the result of our method is the most close to the ground truth. The angle-based method fails to prune the noisy branch in upper right wing, shown in the red rectangle in Fig.~\ref{fig:butterflys:angle}. The SAT method ignores a small noisy branch in the tail, shown in the zoom-in view, while a branch of the right wing has been filtered erroneously as shown in Fig.~\ref{fig:butterflys:sat}, highlighted with a red dash rectangle. Our algorithm keeps the skeleton properly, as shown in the green rectangle in Fig.~\ref{fig:butterflys:dis}. Besides its power in pruning noisy branches, our result is also most accurate by achieving an approximation error $0.1\%$, much less than $0.34\%$, the error of the other two methods.

\subsection{Advantages of spline representation}
As mentioned before, spline curves provide a compact and smooth representation of medial axis transforms. Meanwhile, the control points of spline curves make the medial axis transform much easier to manipulate in applications such as shape editing and deformation.

\begin{figure*}[htbp]
\centering
    \begin{subfigure}[t]{0.49\textwidth}
    \centering
    \includegraphics[scale=0.14]{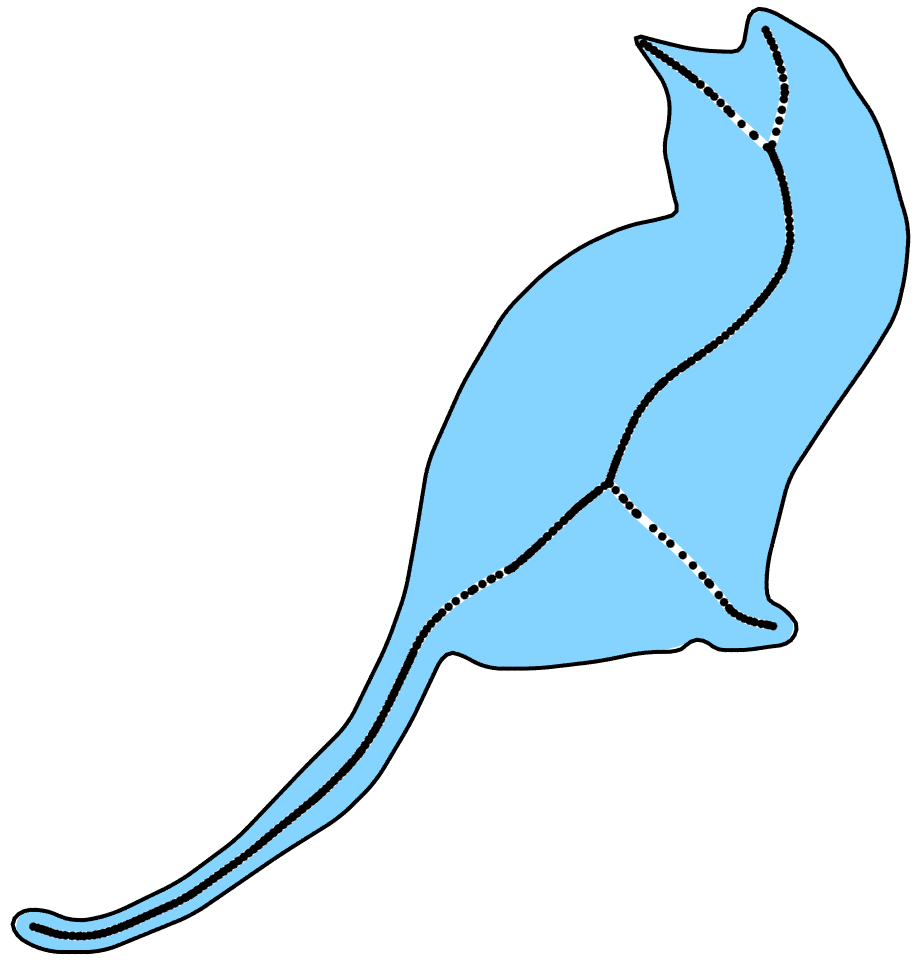}
    \includegraphics[scale=0.14]{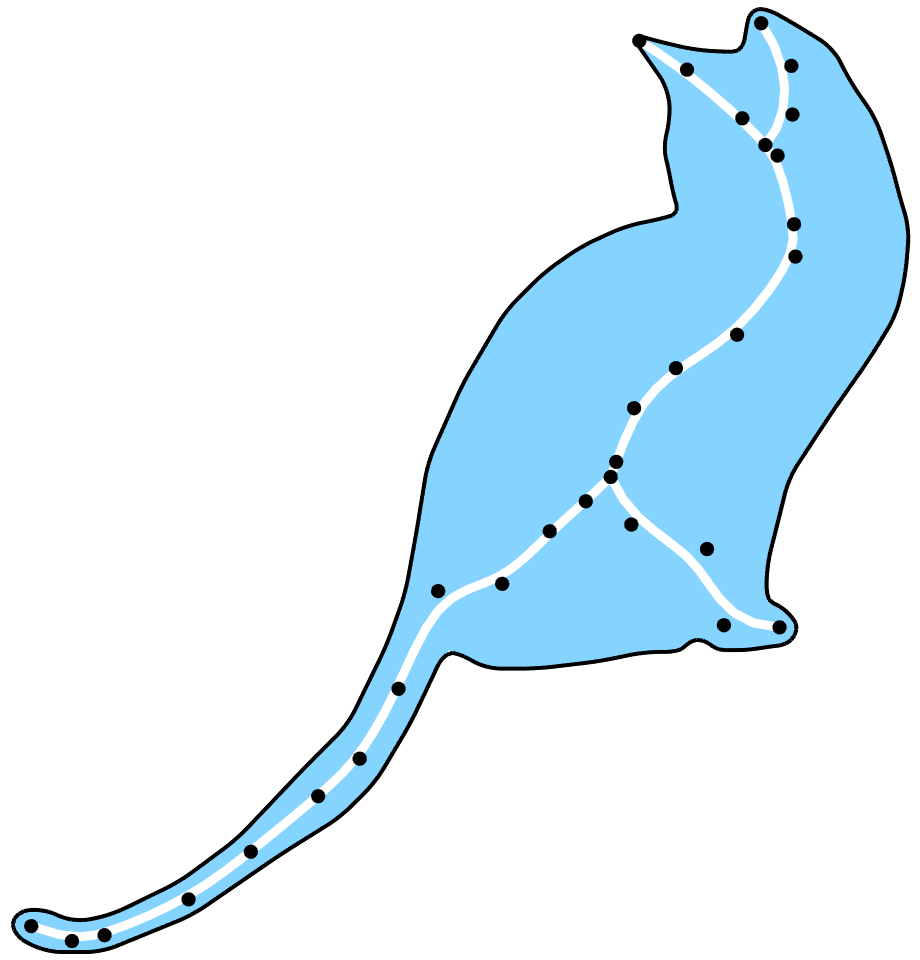}
    \caption{Cat}
    \label{fig:splineadvantages:cat}
    \end{subfigure}
    \hfill
    \begin{subfigure}[t]{0.50\textwidth}
    \centering
    \includegraphics[width=0.49\textwidth]{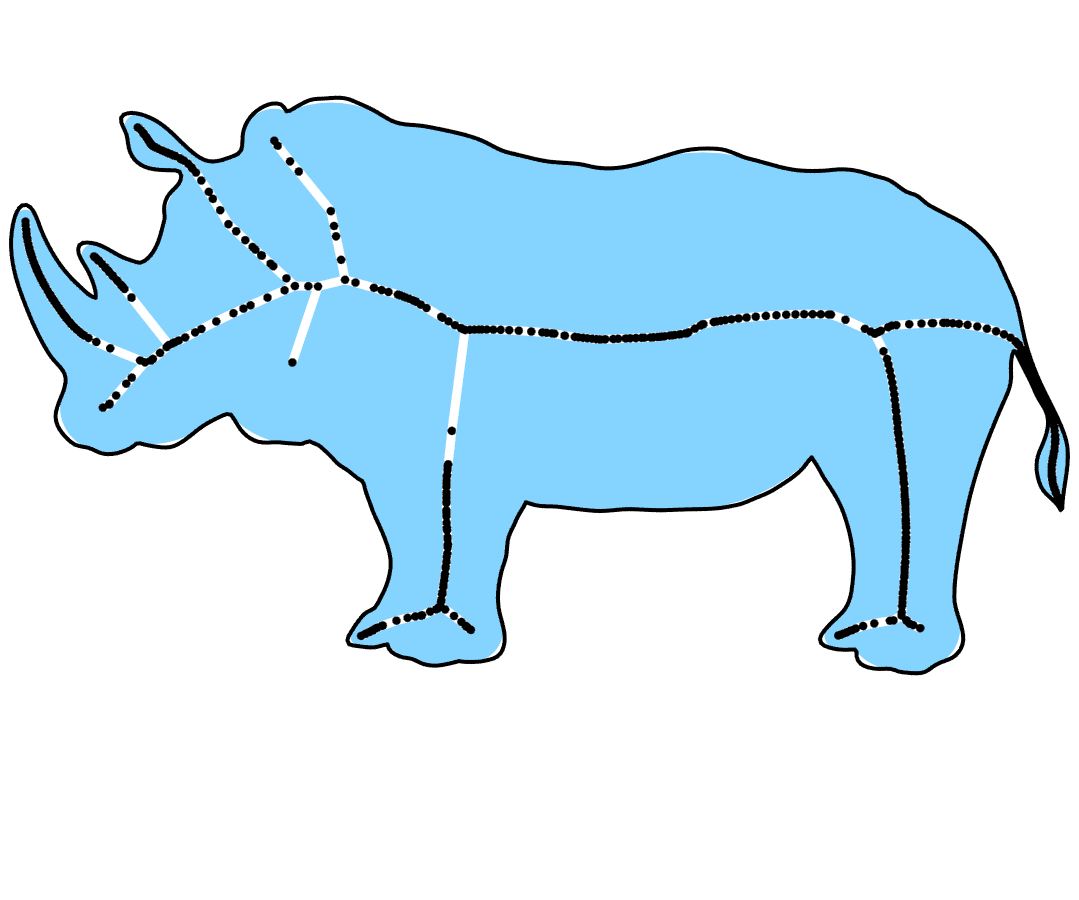}
    \includegraphics[width=0.49\textwidth]{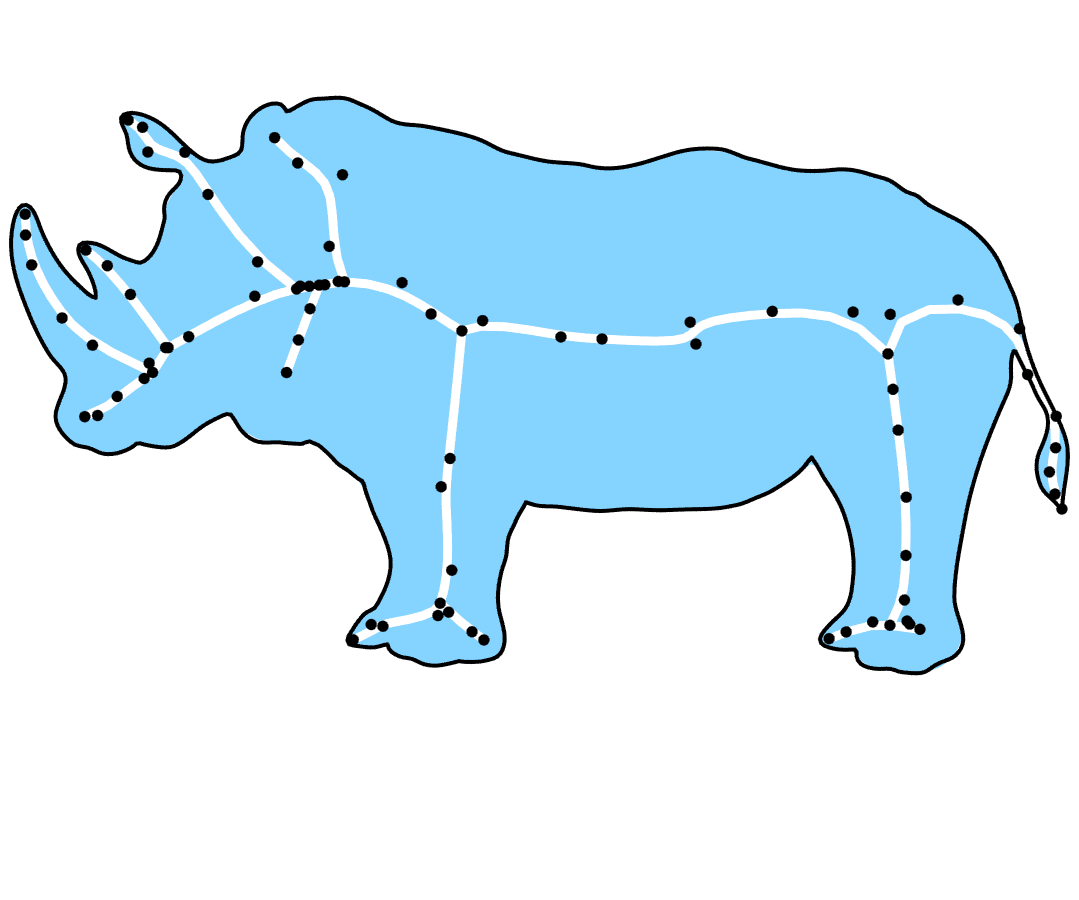}
    \caption{Rhinoceros}
    \label{fig:splineadvantages:rhinoceros}
    \end{subfigure}
    \hfill
\caption{\small Comparisons of discrete medial axis transform and spline medial axis transform. (a) Medial axis transforms of a cat. (b) Medial axis transforms of a rhinoceros. In both comparison sets, the left is a discrete medial axis transform and the right is a piecewise cubic B-spline medial axis transform.}
\label{fig:splineadvantages}
\end{figure*}

Fig.~\ref{fig:splineadvantages} shows comparisons between discrete medial axis transforms and the corresponding piecewise cubic B-spline medial axis transforms.
The number of control points are largely reduced in the spline representation, compared with the number of medial points in the discrete medial axis transform, while the approximation error of the spline representation is comparable with that of the discrete medial axis transform. Table~\ref{tab:stats1} lists the number of medial/control points and approximation errors in our test shapes.

\begin{table}[htbp]
\begin{center}
\footnotesize
\small\addtolength{\tabcolsep}{-4pt}
\renewcommand{\arraystretch}{1.5}
\begin{tabular}{ c | c | c | c | c | c | c }
  \hline
  \multirow{2}{*}{Shape $\cO$}     & \multirow{2}{*}{\texttt{\#}$\bvec{p}_i$} & \multicolumn{2}{|c}{$\varepsilon$} & \multicolumn{2}{|c|}{$|V|$} & \multirow{2}{*}{Compactness}\\ \cline{3-6}
  & & $\varepsilon(\cO, \widehat \cM_s)$ & $\varepsilon(\cO, \widehat \cM)$ & $\cM_s$ & $\cM$ & \\   \hline
  Cat & 500 & 0.28\% & 0.17\% & 420 & 31  & 92.6\%\\   \hline
  Rhinoceros & $\,$ 897 $\,$  &  0.50\%  & 0.36\%    & 549    & $\,$ 81 $\,$    & 85.2\%  \\   \hline
\end{tabular}
\end{center}
\captionof{table}{\small Comparisons of discrete medial axis transforms and spline medial axis transforms (shapes of Fig.~\ref{fig:splineadvantages}). \texttt{\#}$\bvec{p}_i$ is the number of boundary points. $\varepsilon$ is the approximation error of medial reconstruction to $\cO$. $\cM_s$ represents the discrete medial axis transform, and $\cM$ is the corresponding spline medial axis transform. $|V|$ is the number of medial/control points in $\cM_s$/$\cM$. Compactness records the medial points reduction from $\cM_s$ to $\cM$.}
\label{tab:stats1}
\end{table}


\subsection{Performances}
\begin{figure*}[htbp]
\centering
    \begin{subfigure}[t]{0.50\textwidth}
    \centering
    \includegraphics[scale=0.103]{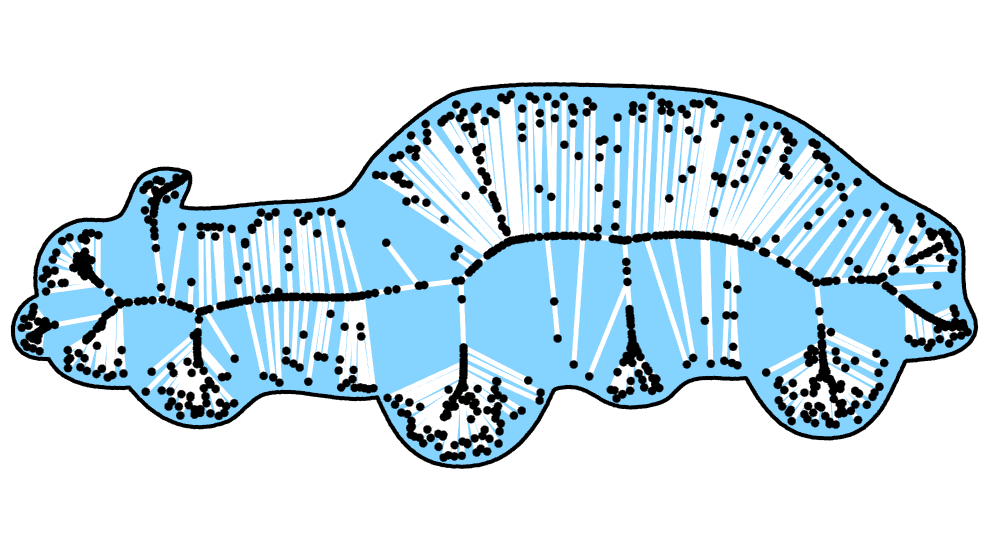}
    \includegraphics[scale=0.103]{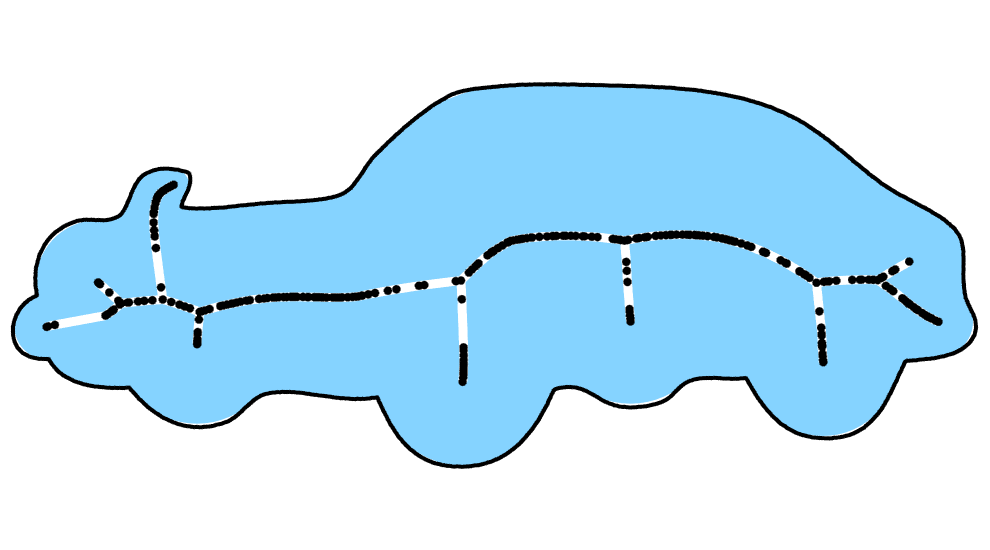}
  	\includegraphics[scale=0.103]{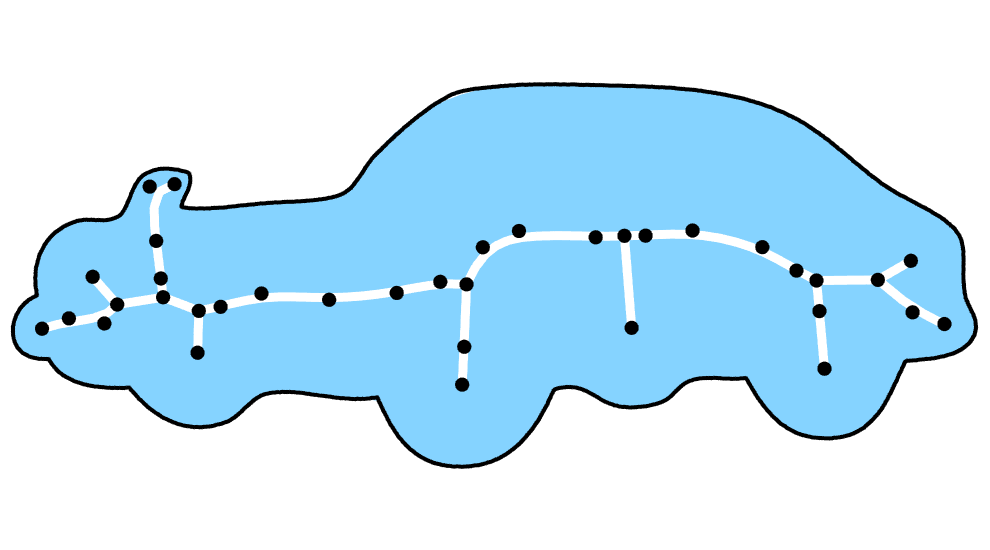}
    \caption{Car}
    \label{fig:performances:car}
    \end{subfigure}
    \hfill
    \begin{subfigure}[t]{0.49\textwidth}
    \centering
    \includegraphics[scale=0.102]{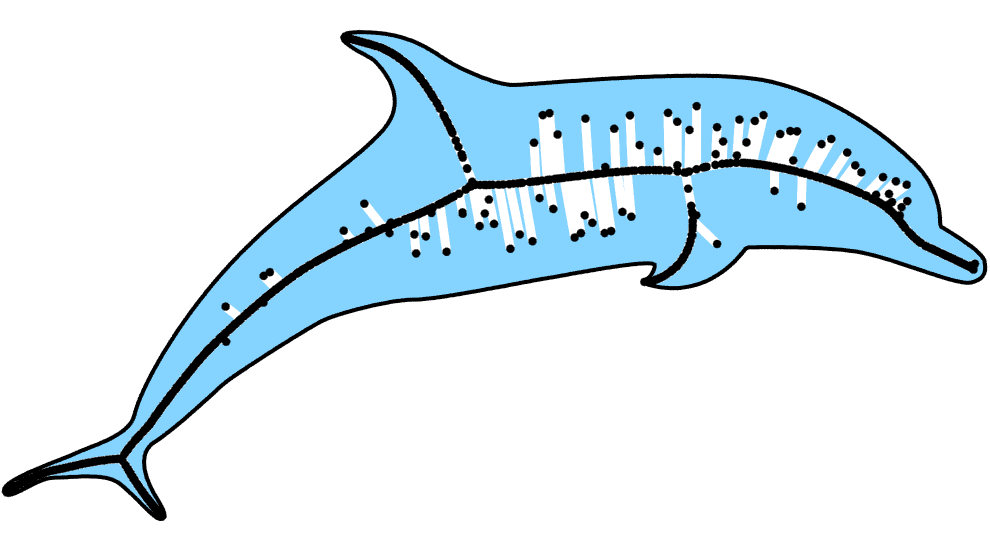}
    \includegraphics[scale=0.102]{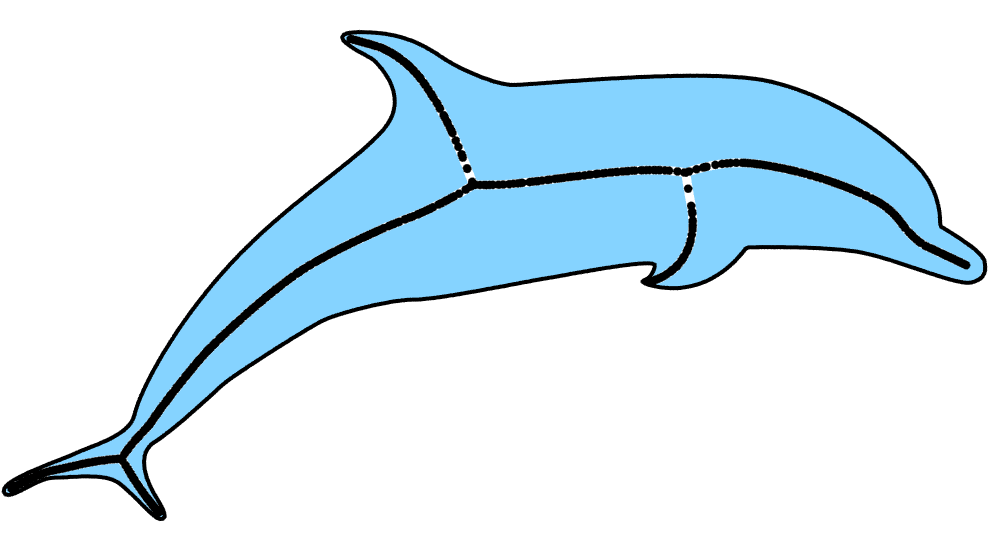}
  	\includegraphics[scale=0.102]{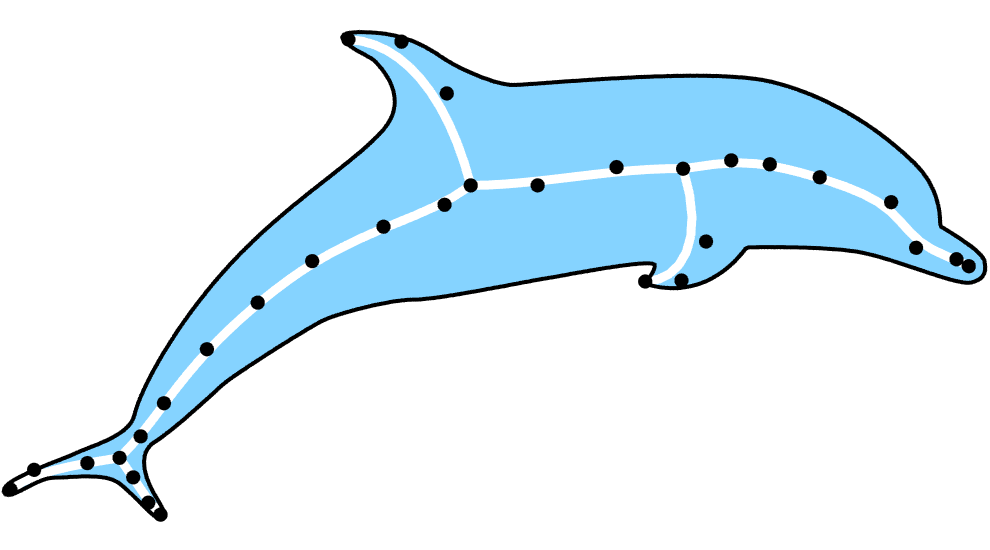}
    \caption{Dolphin}
    \label{fig:performances:dolphin}
    \end{subfigure}
    \hfill
    \begin{subfigure}[t]{0.50\textwidth}
    \centering
    \includegraphics[scale=0.103]{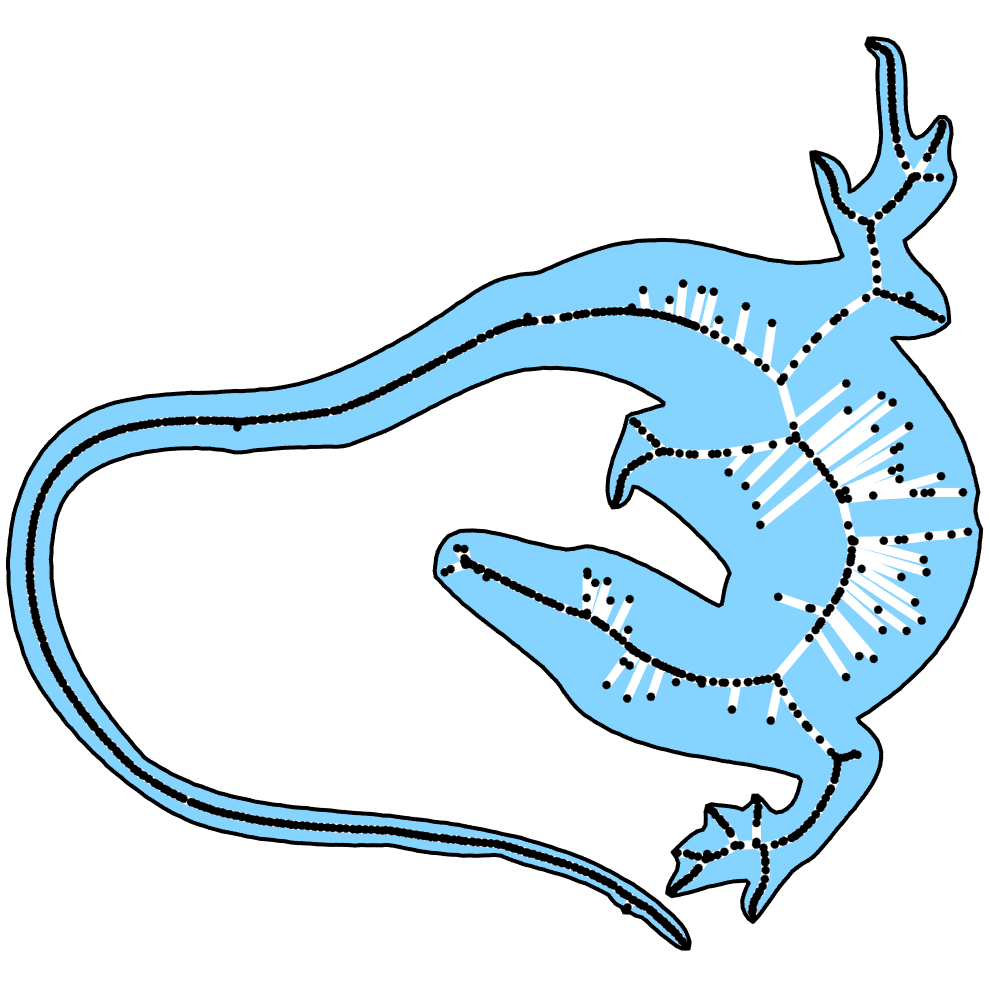}
    \includegraphics[scale=0.103]{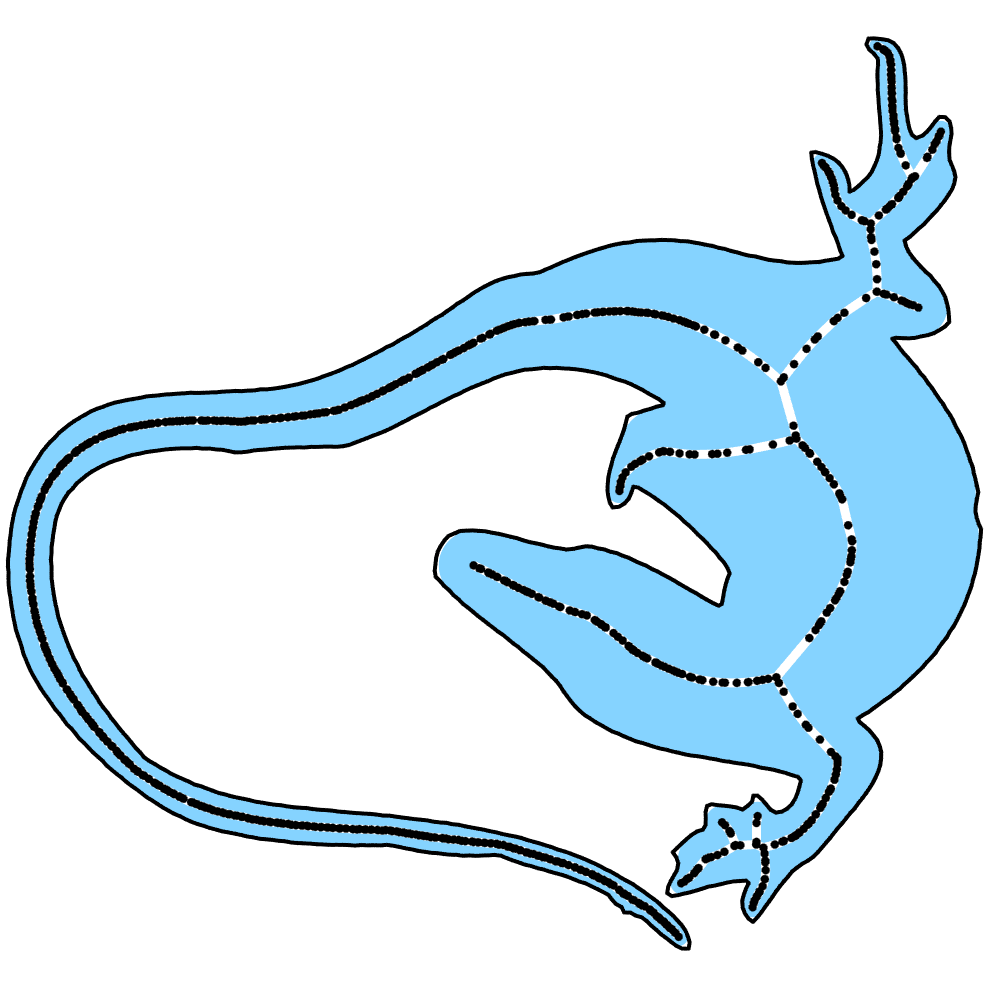}
  	\includegraphics[scale=0.103]{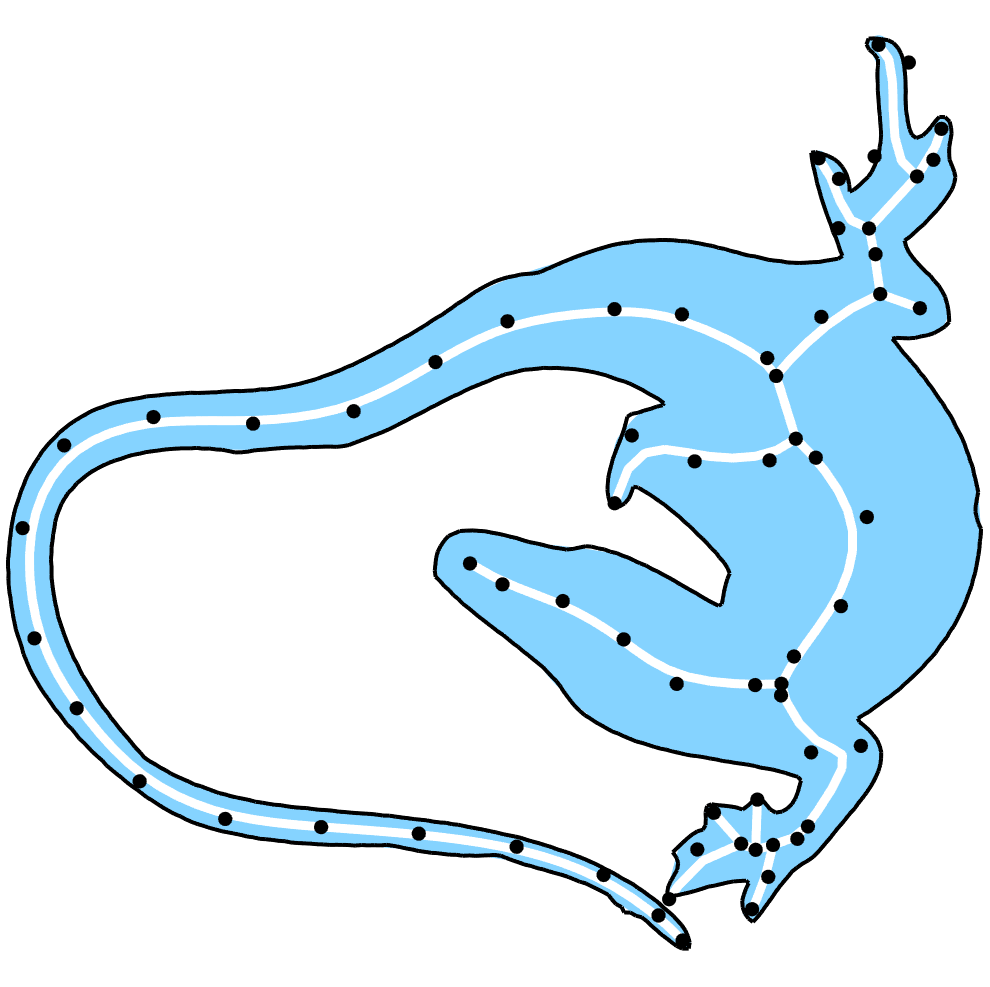}
    \caption{Lizard}
    \label{fig:performances:lizard}
    \end{subfigure}
    \hfill
    \begin{subfigure}[t]{0.49\textwidth}
    \centering
    \includegraphics[scale=0.102]{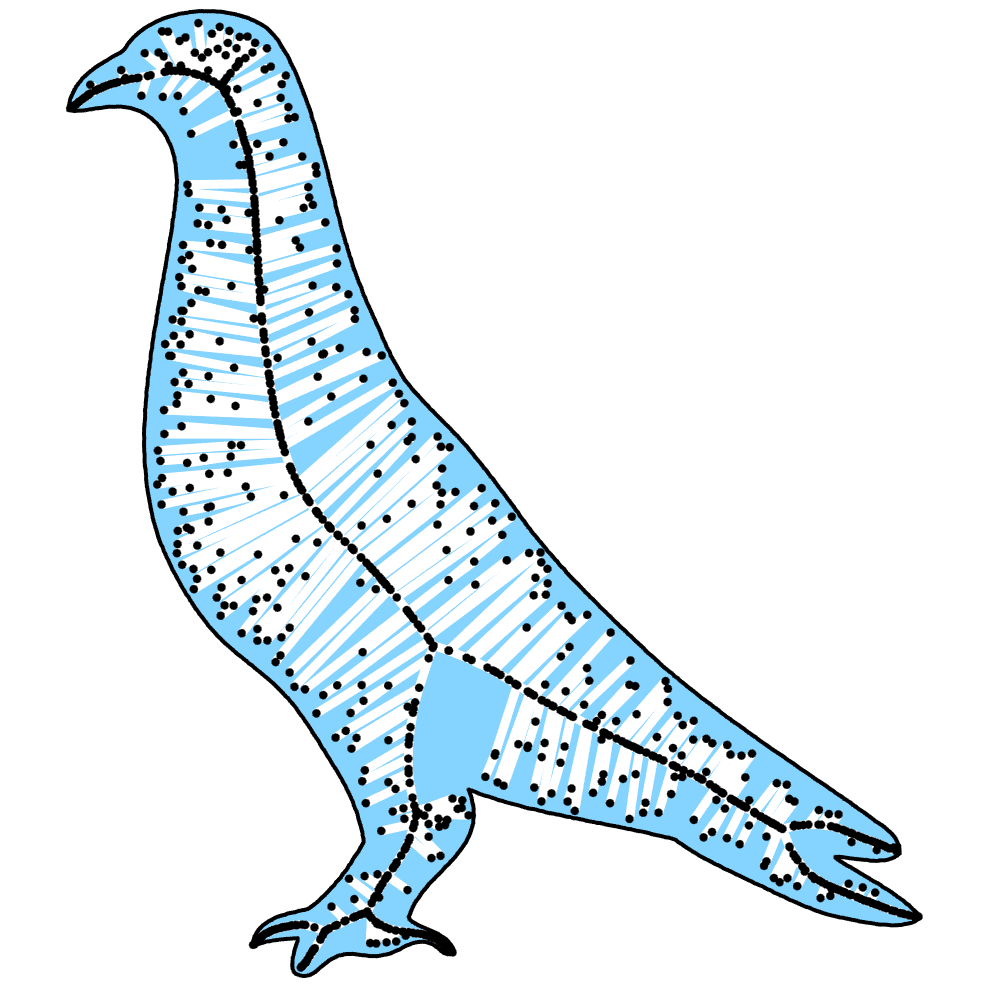}
    \includegraphics[scale=0.102]{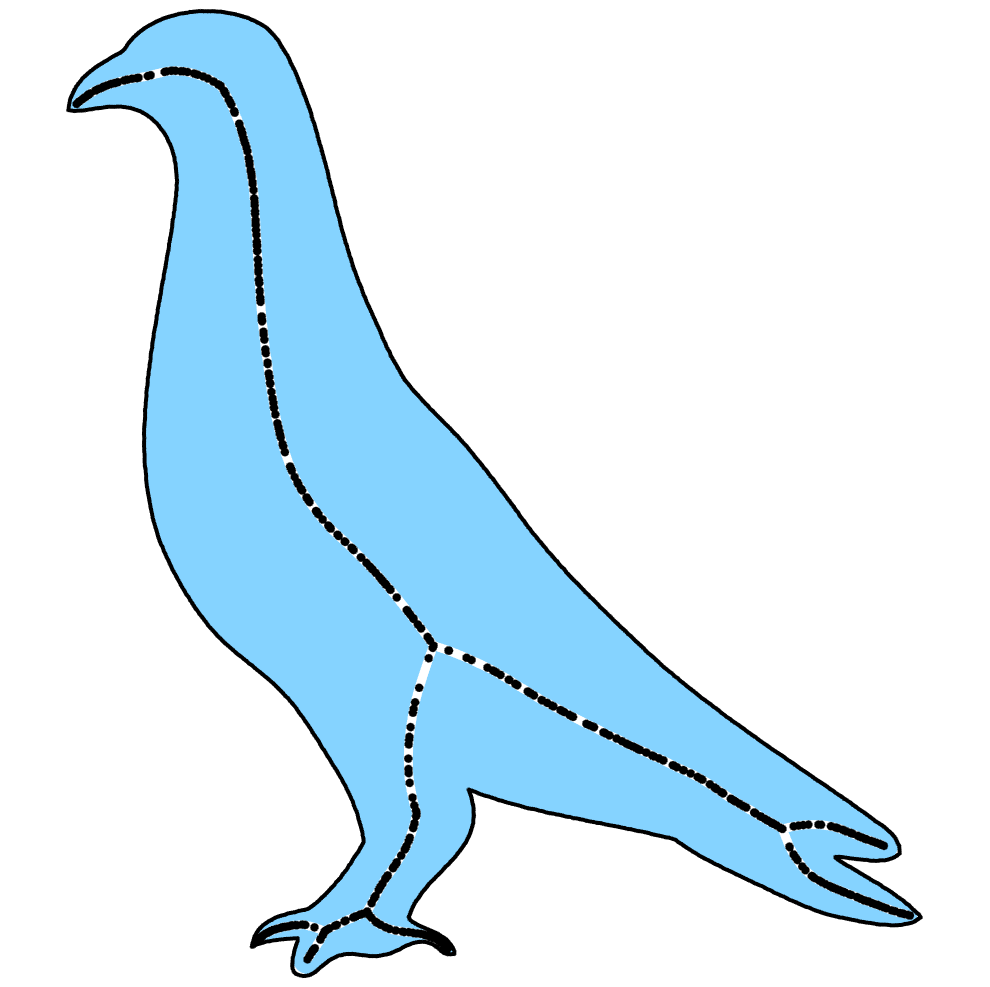}
  	\includegraphics[scale=0.102]{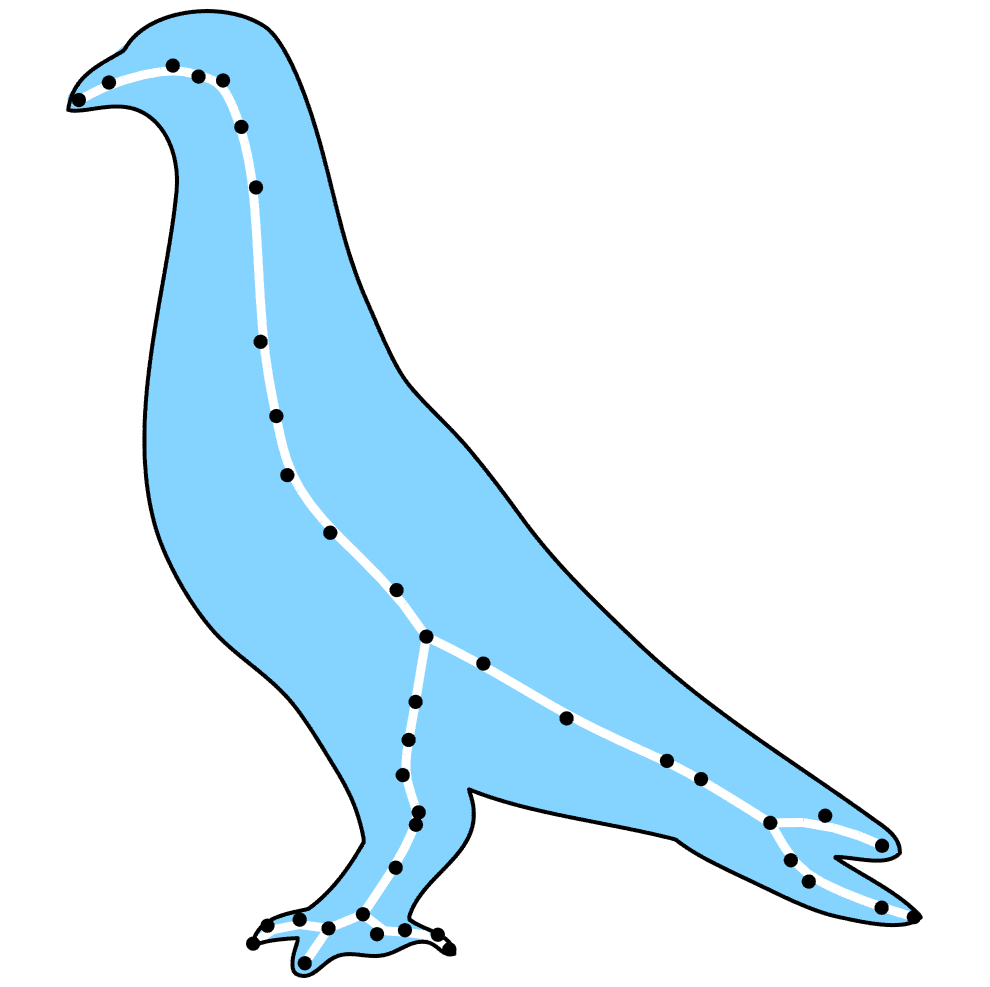}
    \caption{Bird}
    \label{fig:performances:bird}
    \end{subfigure}
    \hfill
    \begin{subfigure}[t]{0.50\textwidth}
    \centering
    \includegraphics[scale=0.103]{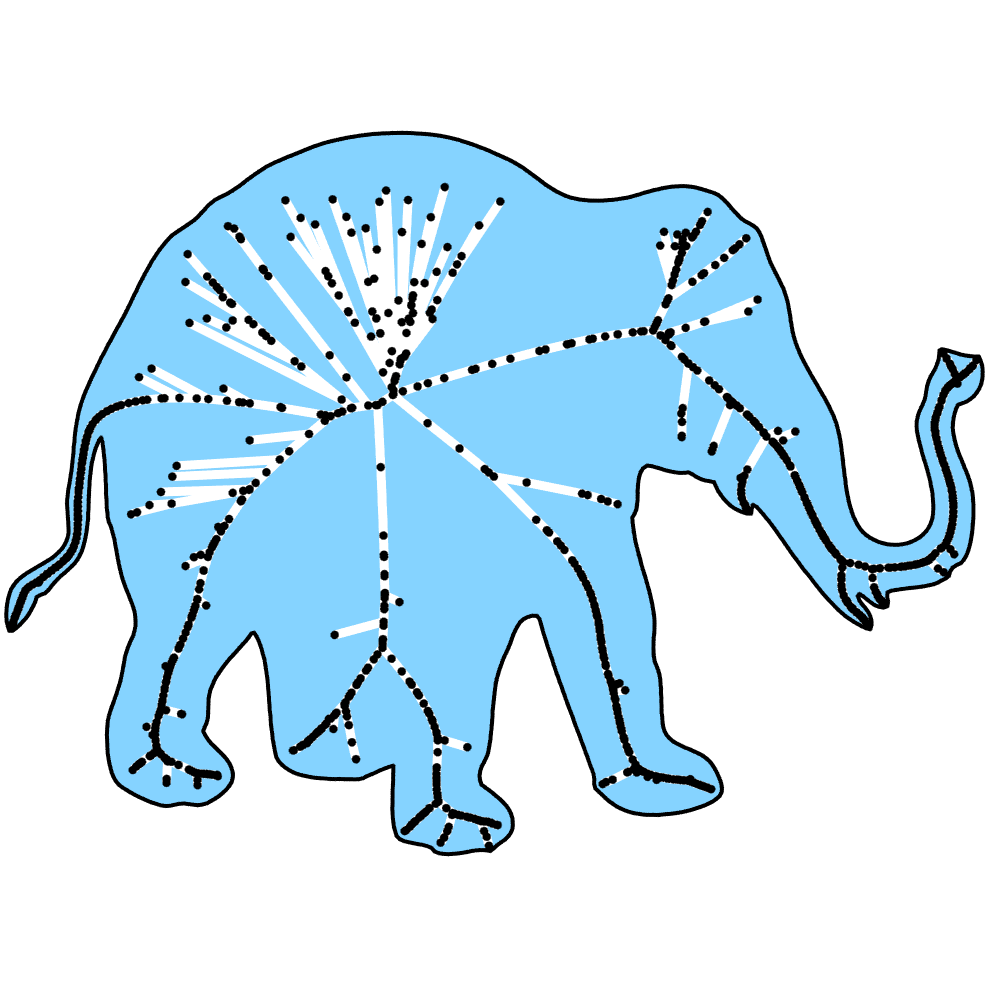}
    \includegraphics[scale=0.103]{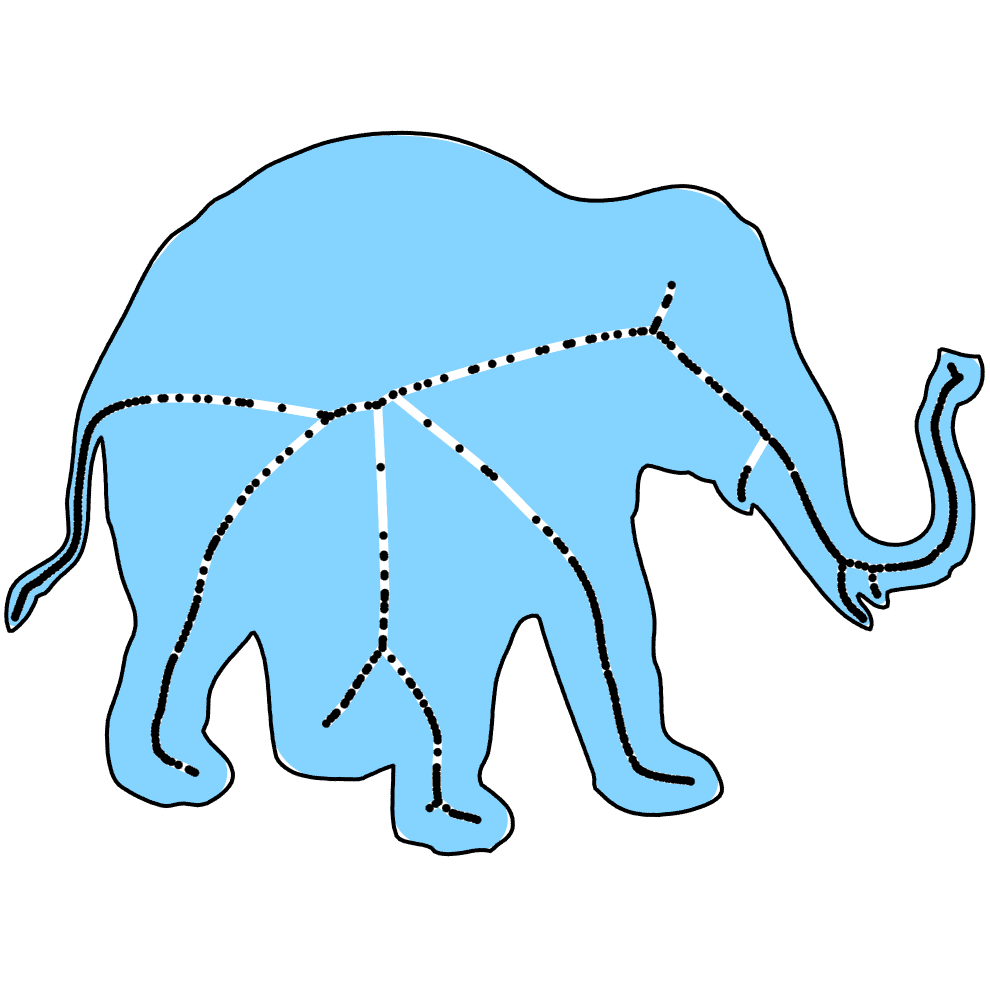}
  	\includegraphics[scale=0.103]{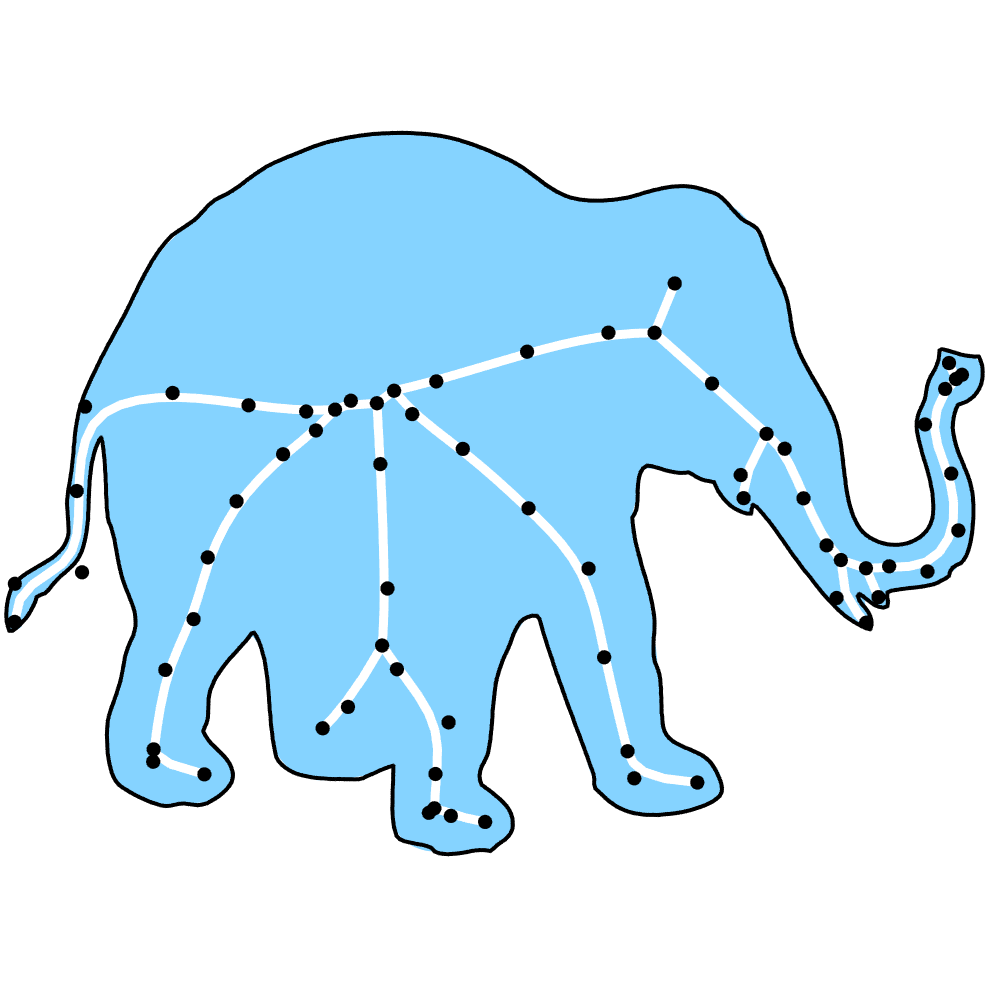}
    \caption{Elephant}
    \label{fig:performances:elephant}
    \end{subfigure}
    \hfill
    \begin{subfigure}[t]{0.49\textwidth}
    \centering
    \includegraphics[scale=0.102]{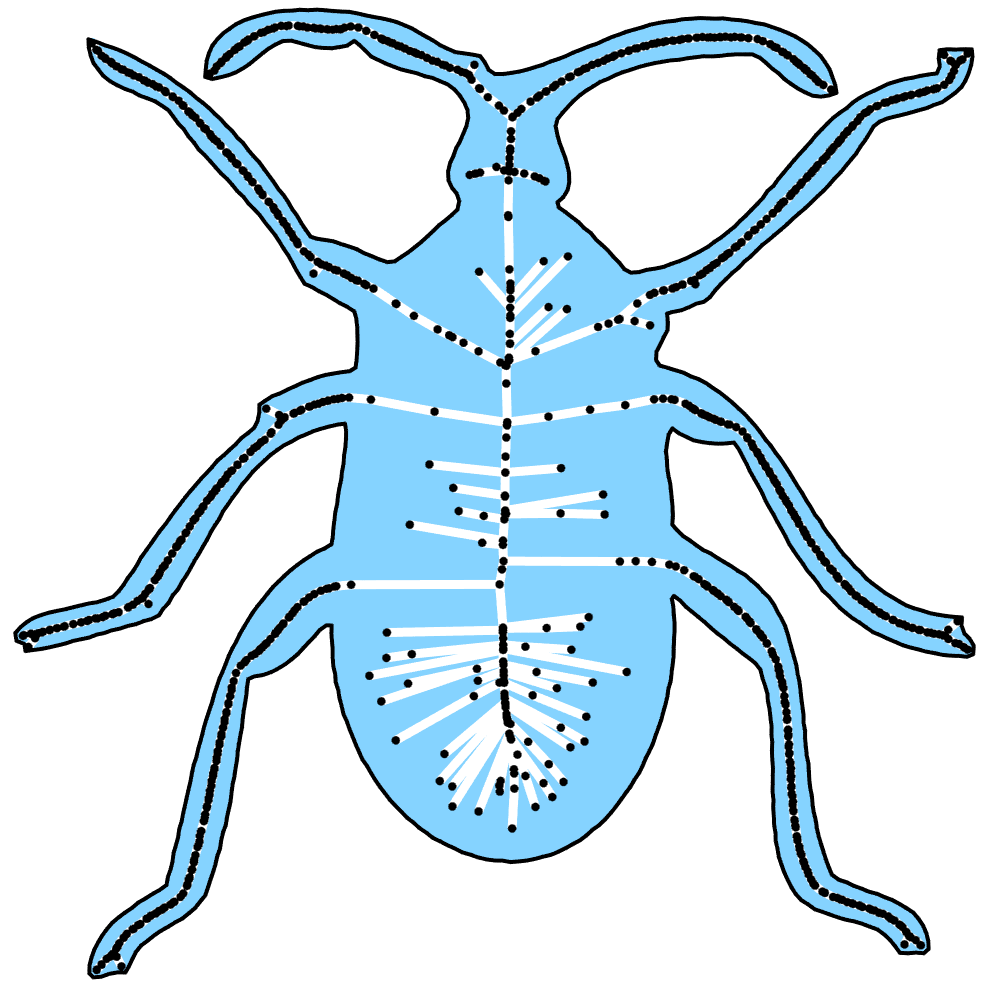}
    \includegraphics[scale=0.102]{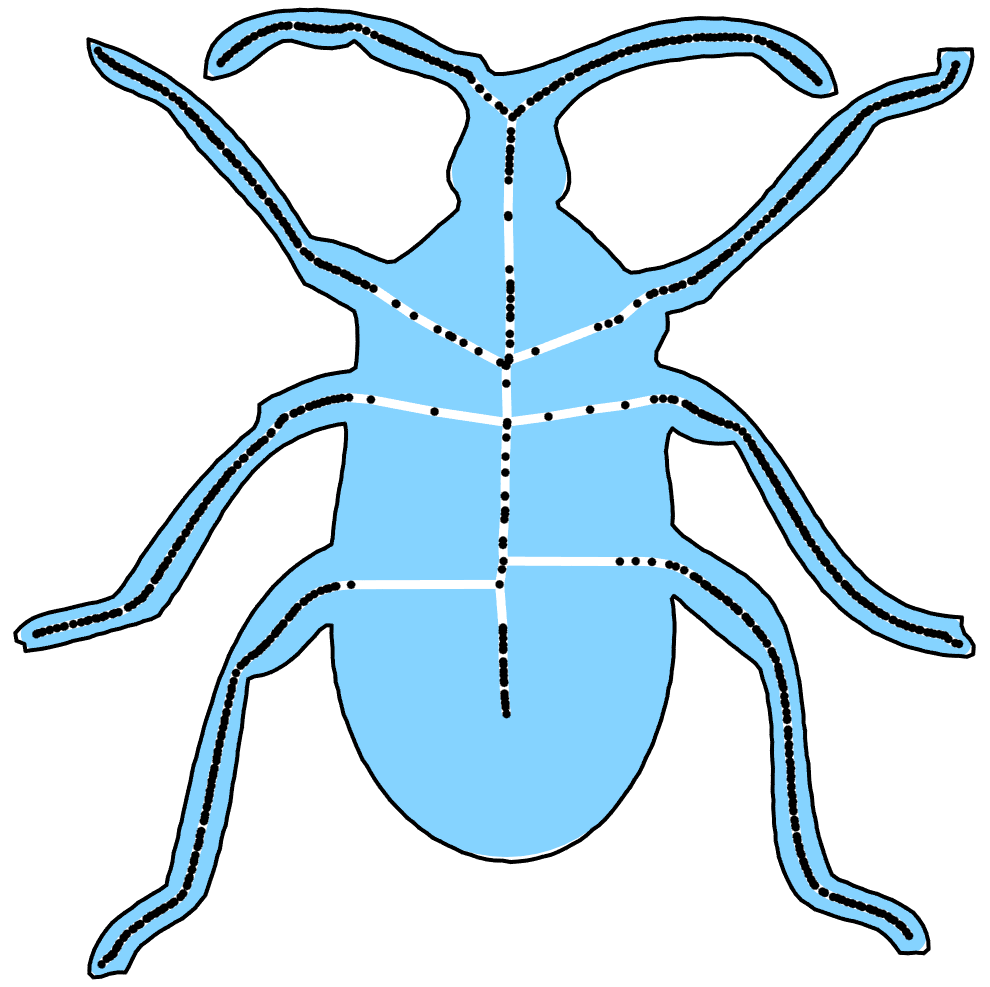}
  	\includegraphics[scale=0.102]{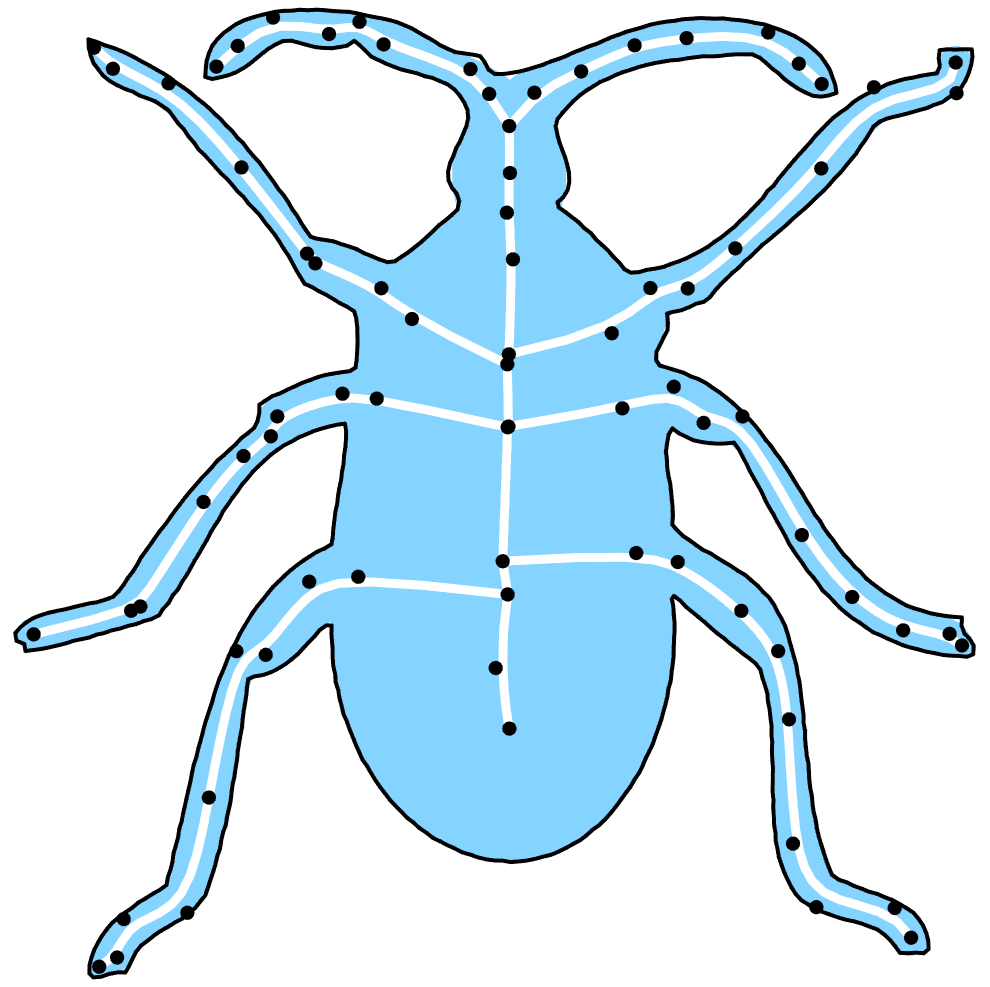}
    \caption{Beetle}
    \label{fig:performances:beetle}
    \end{subfigure}
    \hfill


    \begin{subfigure}[t]{0.49\textwidth}
    \centering
    \includegraphics[scale=0.102]{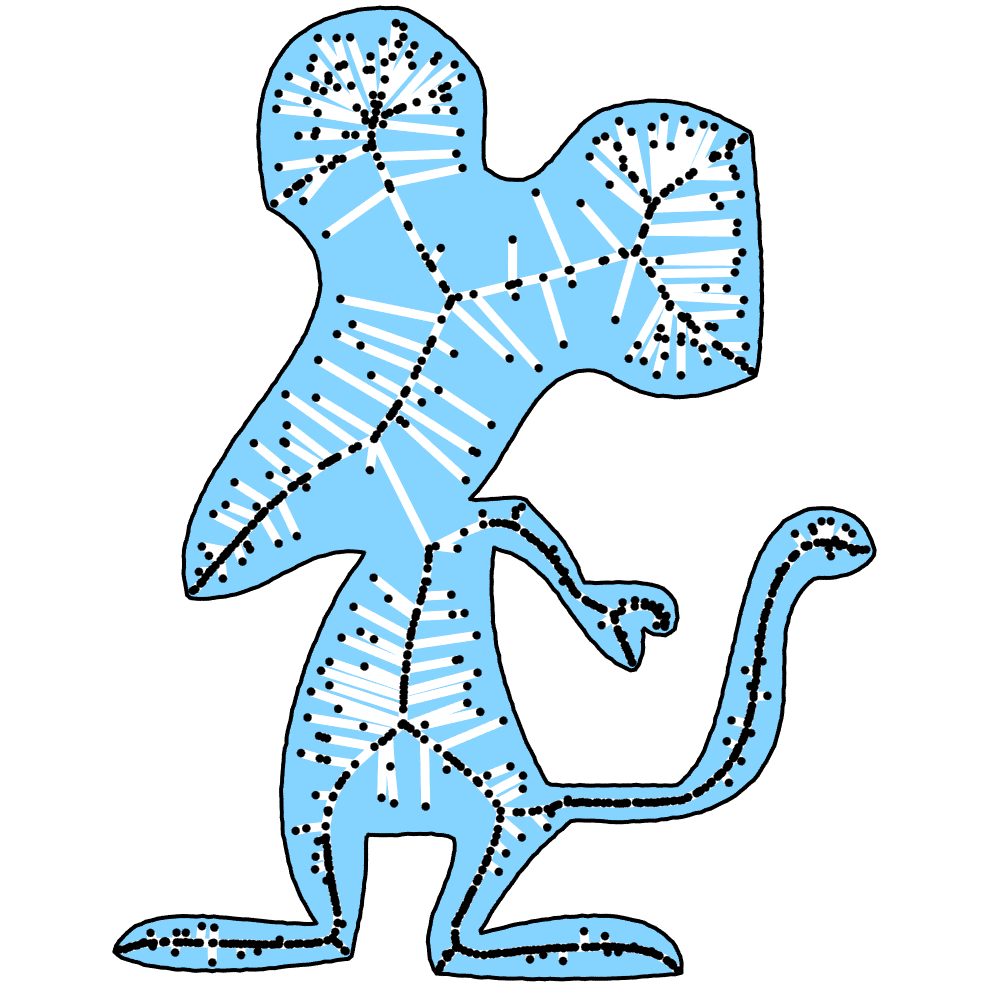}
    \includegraphics[scale=0.102]{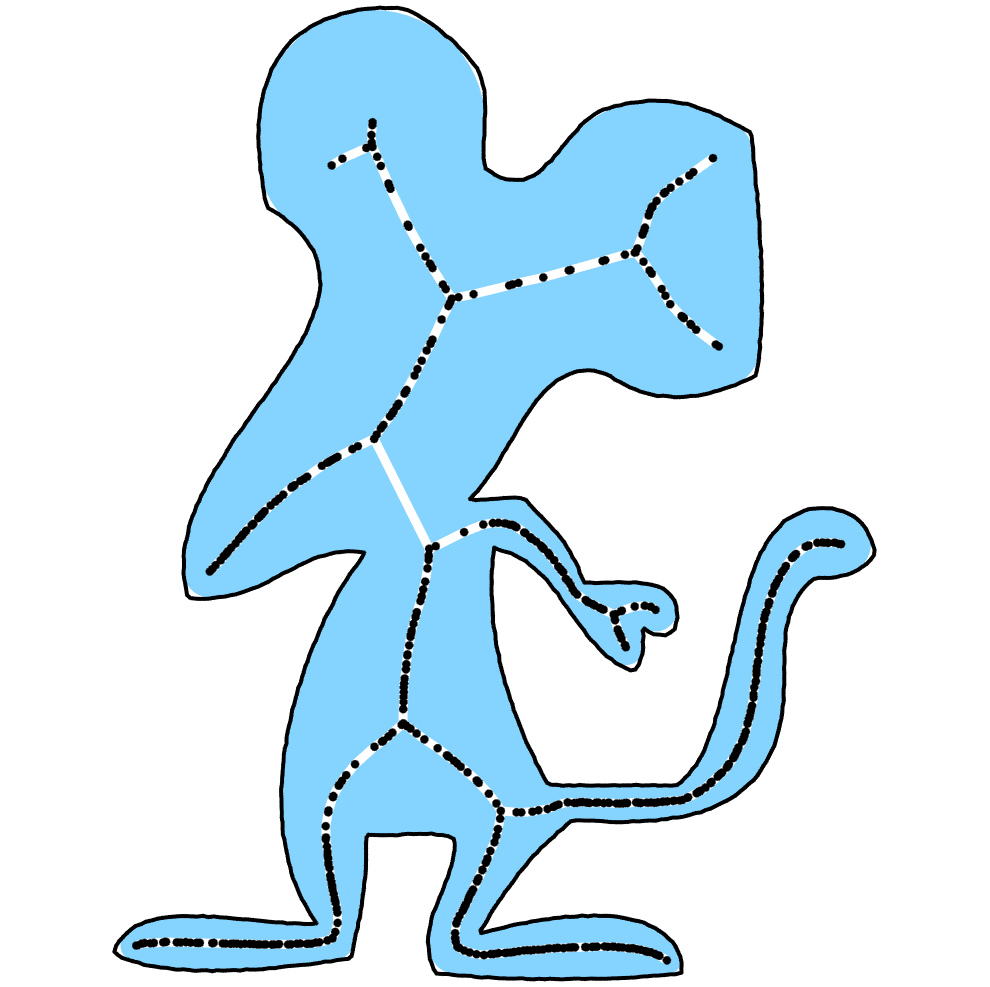}
  	\includegraphics[scale=0.102]{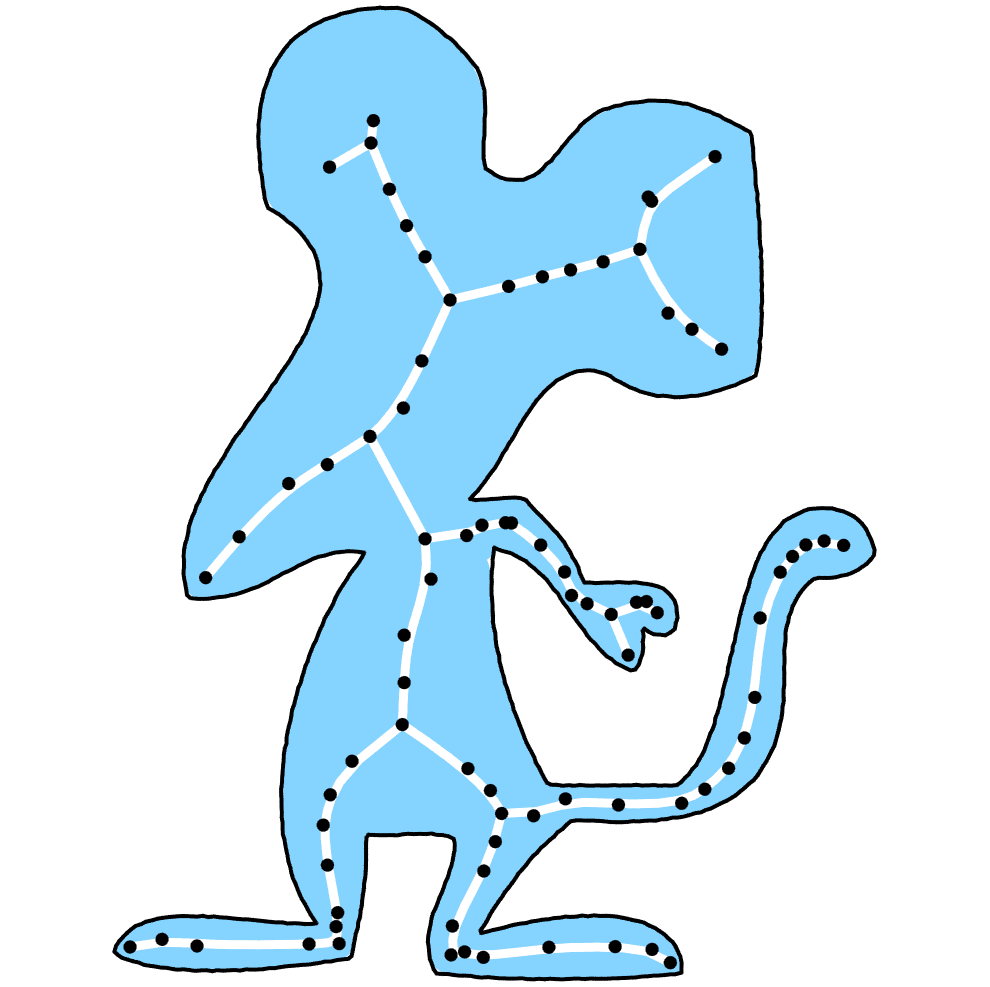}
    \caption{Mouse}
    \label{fig:performances:mouse}
    \end{subfigure}
    \hfill
    \begin{subfigure}[t]{0.50\textwidth}
    \centering
    \includegraphics[scale=0.103]{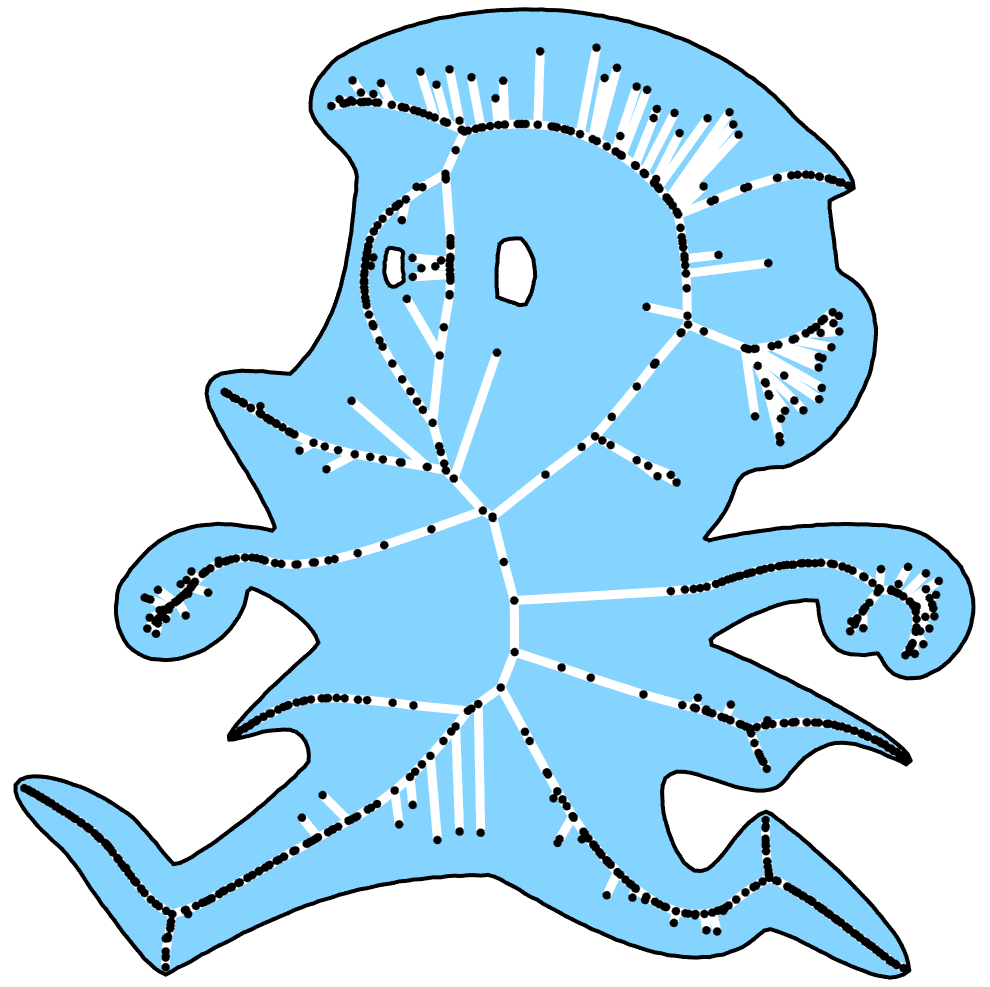}
    \includegraphics[scale=0.103]{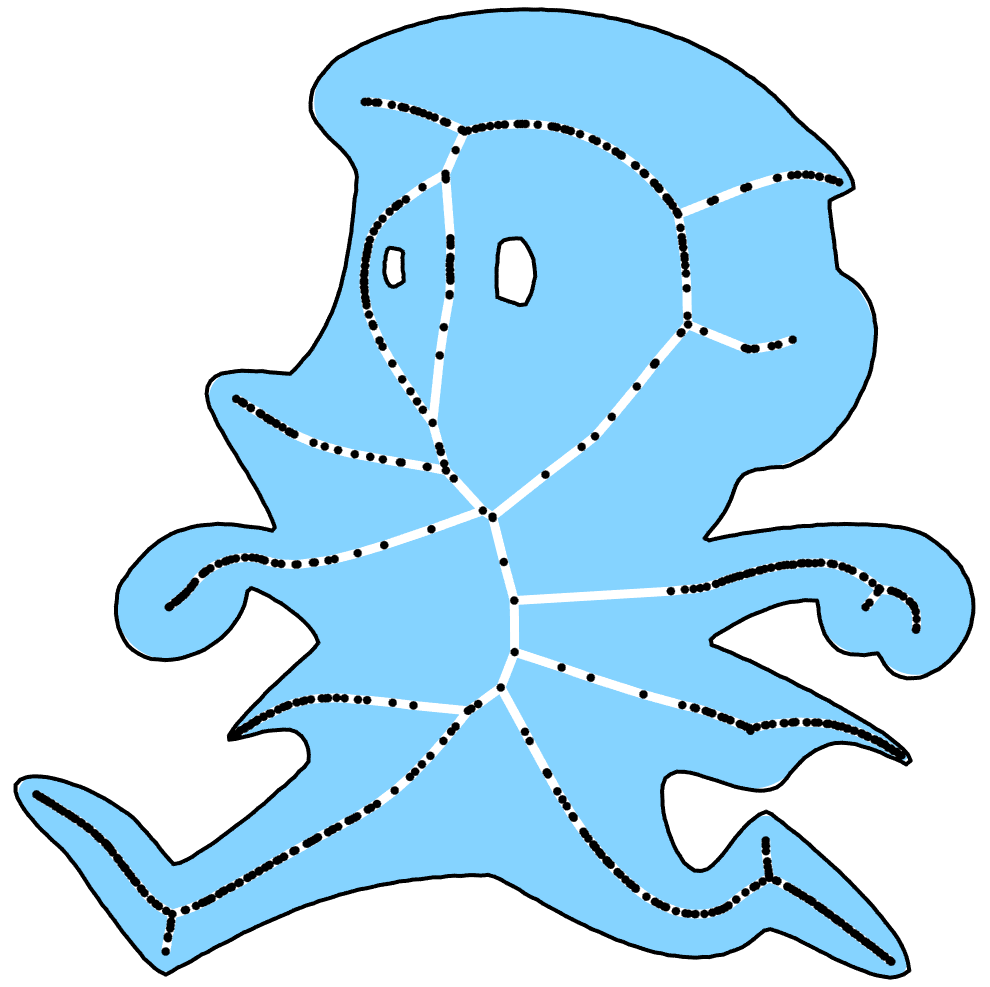}
  	\includegraphics[scale=0.103]{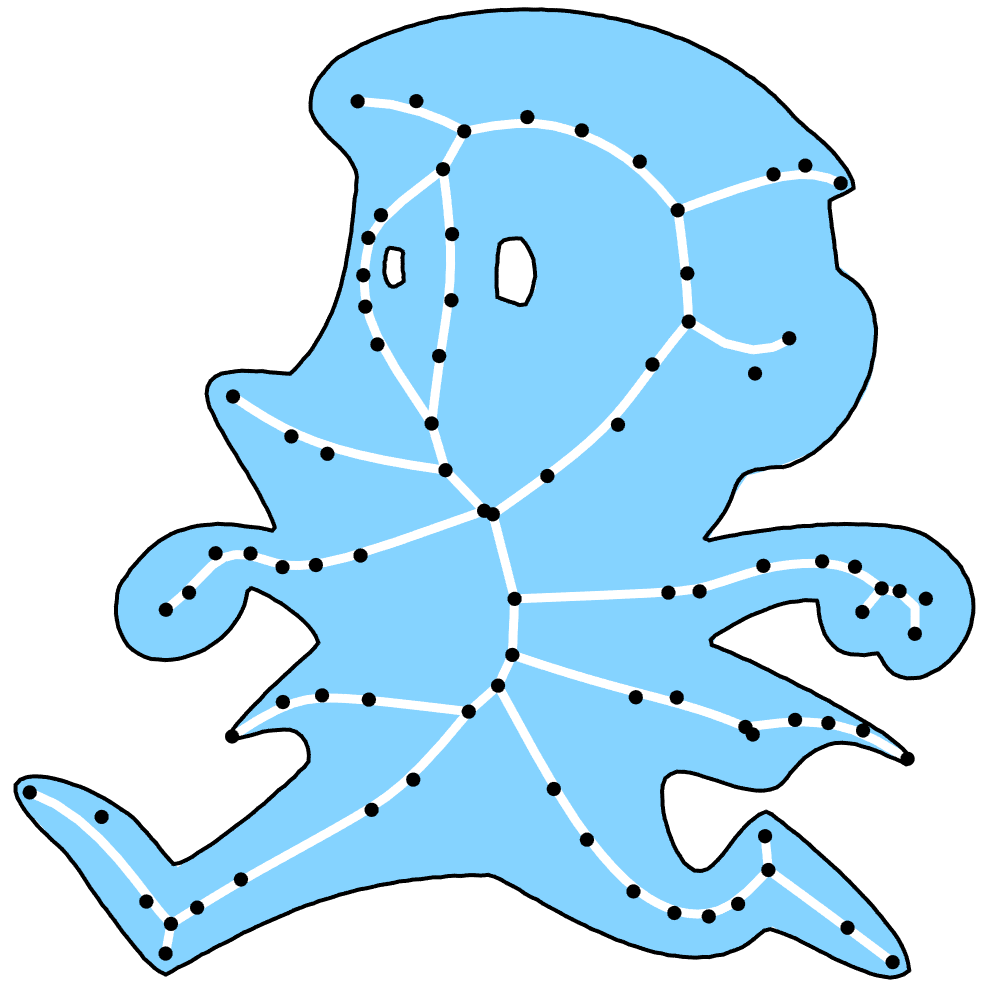}
    \caption{Superman}
    \label{fig:performances:superman}
    \end{subfigure}
\caption{\small Compact spline medial axis transform results on different shapes. In each set of shapes, the left one is the initial medial axis transform $\cM_0$ generated from a Voronoi-based approach; the middle one is a stable medial axis transform $\cM_s$ computed by our noise pruning algorithm; the right one is a piecewise cubic B-spline medial axis transform $\cM$ optimized by our geometric simplification algorithm.}
\label{fig:performances}
\end{figure*}

We tested our algorithm on various 2D shapes. Some of the results are shown in Fig.~\ref{fig:performances}. In each set of shapes, three medial axis transforms are displayed: the initial $\cM_0$ generated from a Voronoi-based approach; a stable medial axis transform $\cM_s$ computed by Algorithm~\ref{alg:topological} and a compact spline-based medial axis transform $\cM$ optimized by Algorithm~\ref{alg:geosimp}.

These examples demonstrate that our algorithm provides an efficient way to obtain faithful and compact medial axis transforms for different 2D shapes, following the control of specified error thresholds.
The statistical analysis in Table~\ref{tab:stats2} shows that the number of points required to approximate the medial axis transform is greatly scaled down with our spline representation of the medial axis transform. In addition, the L-BFGS method assures the efficiency of our optimization process.

\begin{table*}[htbp]
\begin{center}
\footnotesize
\small\addtolength{\tabcolsep}{-4pt}
\renewcommand{\arraystretch}{1.5}
\begin{tabular}{ c | c | c | c | c | c | c | c | c | c}
  \hline
  \multirow{2}{*}{Shape $\cO$}     & \multirow{2}{*}{\texttt{\#}$\bvec{p}_i$} & \multirow{2}{*}{$\hat \varepsilon$} & \multicolumn{2}{|c}{$\varepsilon$} & \multicolumn{3}{|c|}{$|V|$} & Compactness & Time(sec)  \\ \cline{4-10}
  & & & $\,$ $\varepsilon(\cO, \widehat \cM_s)$ $\,$ & $\,$ $\varepsilon(\cO, \widehat \cM)$ $\,$ & $\cM_0$ & $\cM_s$ & $\cM$  & \multicolumn{2}{|c}{$\cM_s \rightarrow \cM$} \\ \hline
  Car & 1000 & 0.40\% & 0.40\% & 0.32\% & 998 & 295 & 36 & 87.8\% & 0.08 \\ \hline
  Dolphin & 800 & 0.25\% & 0.24\% & 0.15\% & 798 & 685 & 31 & 95.5\% & 0.08 \\ \hline
  Lizard & $\quad$ 1000 $\quad$  & $\quad$ 0.50\% $\quad$  & $\,$ 0.49\% $\,$     & $\,$ 0.48\% $\,$   & $\quad$ 998 $\quad$ & $\quad$ 836 $\quad$    & $\quad$ 65 $\quad$    & 92.2\% & 0.11  \\ \hline
  Bird & 1000 & 0.30\% & 0.30\% & 0.28\% & 998 & 559 & 37 & 93.4\% & 0.13 \\ \hline
  Elephant & 1000 & 0.50\% & 0.50\% & 0.44\% & 998 & 660 & 67 & 89.8\% & 0.15 \\ \hline
  Beetle & 1000 & 0.50\% & 0.50\% & 0.43\% & 998 & 872 & 79 & 90.9\% & 0.18 \\ \hline
  Mouse & 1000 & 0.50\% & 0.50\% & 0.34\% & 998 & 619 & 81 & 86.9\% & 0.26 \\ \hline
  Superman & 851 & 0.35\% & 0.34\% & 0.32\% & 853 & 633 & 85 & 86.6\% & 0.40 \\ \hline
\end{tabular}
\end{center}
\captionof{table}{\small Performances of our algorithm (shapes of Fig.~\ref{fig:performances}). \texttt{\#}$\bvec{p}_i$ is the number of boundary points. $\hat \varepsilon$ is the user-definded error threshold. $\varepsilon$ is the approximation error of medial reconstruction to $\cO$. $|V|$ is the number of medial/control points in medial axes. $\cM_0$ is the initial medial axis transform based on a Voronoi-based approach. $\cM_s$ represents the stable medial axis transform computed by our noise pruning algorithm, and $\cM$ is the spline medial axis transform optimized by our geometric simplification algorithm. Compactness shows the medial points reduction from $\cM_s$ to $\cM$. Time column records the time cost of geometric simplification from $\cM_s$ to $\cM$. }
\label{tab:stats2}
\end{table*}

\section{Conclusion} \label{sec:conclusion}
We propose a framework for computing an accurate and compact medial axis transform for an arbitrary 2D shape. The approximation quality of the medial axis transform is guaranteed to be less than a user specified threshold. Moreover, users have the freedom to choose different types of splines to represent the medial axis transform according to their requirements and applications. A noise pruning algorithm for the medial axis transform is also proposed and integrated in the framework. This noise pruning algorithm filters noise in the medial axis transform robustly, and provides a good initial medial axis transform for optimization in our framework. Experimental results confirm the effectiveness of our algorithm.
In future, we plan to study how to apply the new spline-based representation of the medial axis transform in shape deformation and shape matching, and extend the framework for the computation of the medial axis transform of a 3D volume.


\appendix
\section{Error term computation} \label{sec:app:errorterm}
As discussed in Section~\ref{sec:geosimp}, the energy function $E(\bvec X)$ (Equation~\ref{eq:2dmatenergy}) could be computed from $d^2(\bvec{p}_i,  \partial \widehat S)$, which is the squared Euclidean distance from
boundary point $\bvec{p}_i$ to the envelope boundary of a segment $S$ sampled on the curves of $\cM$.
For each segment $S$, there are two points $\bvec{v}_1$ and $\bvec{v}_2$, which represent two medial circles $(\bvec{u}_1, r_1)$ and $(\bvec{u}_1, r_1)$ in 2D space.
And there are one or two common external tangent for two medial circles, which is implied by the property of the medial axis transform.

\begin{figure}[htbp]
\centering
  \includegraphics[width=0.8\linewidth]{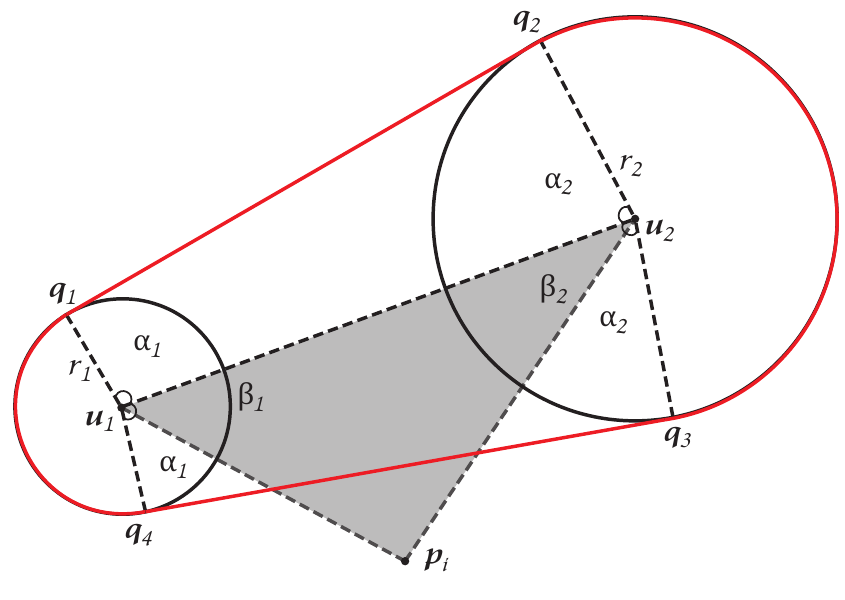}
\caption{\small The envelope of two medial circles $(\bvec{u}_1, r_1)$ and $(\bvec{u}_2, r_2)$.}
\label{fig:envelope}
\end{figure}

Consider the case where two circles have two distinct external tangent lines (Fig.~\ref{fig:envelope}).
Let ${\bvec{q}_i ( i = 1, 2, 3, 4)}$ be the tangent points. The external tangent lines through $\bvec{q}_i$ are represented as $\mc{L}_{12}$ and $\mc{L}_{34}$.
The circle is partitioned into two circular arcs by tangent points, and the two outer arcs are denoted as $\mc{A}_{14}$ and $\mc{A}_{23}$, respectively. The boundary of the shape represented by $S$, denoted by $\partial \widehat S$, consists of two outer arcs $\mc{A}_{14}$, $\mc{A}_{23}$ and two external tangent lines $\mc{L}_{12}$, $\mc{L}_{34}$. Let $\beta_1 = \angle \bvec{p}_i\bvec{u}_1\bvec{u}_2$, $\beta_2 = \angle \bvec{p}_i\bvec{u}_2\bvec{u}_1$, $\alpha _1$ represents $\angle \bvec{q}_1\bvec{u}_1\bvec{u}_2$ or $\angle \bvec{q}_4\bvec{u}_1\bvec{u}_2$, $\alpha _2$ be $\angle \bvec{q}_2\bvec{u}_2\bvec{u}_1$ or $\angle \bvec{q}_3\bvec{u}_2\bvec{u}_1$. By comparing these angles, we can easily tell which part $\bvec{p}_i$'s footpoint lies in. In sequence,
the distance from $\bvec{p}_i$ to its footpoint on $\partial \widehat S$ is computed as
\begin{displaymath}
\tiny
d^2(\bvec{p}_i, \partial \widehat S)=\left
\{ \begin {array}{ll}
( r_1 - \|\bvec{p}_i\bvec{u}_1\|_2)^2 & {\textrm{if } \beta_1 \geq \alpha_1} \\
( r_2 - \|\bvec{p}_i\bvec{u}_2\|_2)^2 & {\textrm{if } \beta_2 \geq \alpha_2}\\
d^2(\bvec{p}_i, \mc{L}_{34}) & {\textrm{if } \beta_1 < \alpha_1 \textrm{ and } \beta_2 < \alpha_2 \textrm{ and} }\\
                          & {\bvec{p}_i, \bvec{u}_1, \bvec{u}_2 \textrm{ are in clockwise order}} \\
d^2(\bvec{p}_i, \mc{L}_{12}) & {\textrm{otherwise}}
\end {array}, \right.
\end{displaymath}
where $d^2(\bvec{p}_i, \mc{L}_{12})$ and $d^2(\bvec{p}_i, \mc{L}_{34})$ are easy to compute, since we know the algebraic representation of two outer tangents, and the gradient of $d^2(\bvec{p}_i, \partial \widehat S)$ can be obtained similarly.

It is also possible that only one common external tangent exists for two medial circles, which occurs when the slope of two medial points equals 1, computed as Equation~\ref{eq:slope}.
In that case, $S$ is a circle and computation of $d^2(\bvec{p}_i, \partial \widehat S)$ is trivial.

\section*{References}
\bibliographystyle{elsarticle-num}
\bibliography{gmpMedialAxis2d-num.v3.11-modify_after_thesis_submission}



%
%
%
\end{document}